\RequirePackage{fix-cm}
\documentclass[twocolumn]{svjour3}          
\smartqed  
\usepackage{atlasphysics}

\usepackage{subfigure}
\usepackage{amsmath,graphicx}

\usepackage{lineno}
\usepackage{multirow,xspace}

\usepackage{shortcuts} 

\usepackage{preprintcover}

\PreprintCoverPaperTitle{Identification and energy calibration of hadronically decaying tau leptons with the ATLAS experiment in $pp$ collisions at $\sqrt{s}$=8~TeV}

\PreprintIdNumber{CERN-PH-EP-2014-227}  

\PreprintCoverAbstract{This paper describes the trigger and offline reconstruction, identification and energy calibration algorithms for hadronic decays of tau leptons employed for the data collected from $pp$ collisions in 2012 with the ATLAS detector at the LHC center-of-mass energy $\sqrt{s}$ = 8 TeV. The performance of these algorithms is measured in most cases with $Z$ decays to tau leptons using the full 2012 dataset, corresponding to an integrated luminosity of 20.3 fb$^{-1}$. An uncertainty on the offline reconstructed tau energy scale of $2-4$\%, depending on transverse energy and pseudorapidity, is achieved using two independent methods. The offline tau identification efficiency is measured with a precision of 2.5\% for hadronically decaying tau leptons with one associated track, and of 4\% for the case of three associated tracks, inclusive in pseudorapidity and for a visible transverse energy greater than $20$ GeV. For hadronic tau lepton decays selected by offline algorithms, the tau trigger identification efficiency is measured with a precision of $2-8$\%, depending on the transverse energy. The performance of the tau algorithms, both offline and at the trigger level, is found to be stable with respect to the number of concurrent proton-proton interactions and has supported a variety of physics results using hadronically decaying tau leptons at ATLAS.}

\PreprintJournalName{Eur. Phys. J. C}

\begin{document}

\sloppy

\title{Identification and energy calibration of hadronically decaying tau leptons with the ATLAS experiment 
in {\bf $pp$} collisions at {\bf $\sqrt{s}$}=8 \TeV{}
}


\author{The ATLAS Collaboration}


%

\authorrunning{The ATLAS Collaboration} 
\institute{ }

\date{Received: date / Accepted: date}

\titlerunning{Identification and energy calibration of hadronically decaying tau leptons in ATLAS at {\bf $\sqrt{s}$}=8 \TeV{}}

\maketitle

\begin{abstract}
This paper describes the trigger and offline reconstruction, identification and energy calibration algorithms for hadronic decays of tau leptons employed for the 
data collected from $pp$ collisions in 2012 with the ATLAS detector at the LHC center-of-mass energy $\sqrt{\mathrm{s}} = 8$\,\TeV . 
The performance of these algorithms is measured in most cases with \Zboson\ decays to tau leptons using the full 2012 dataset, corresponding 
to an integrated luminosity of 20.3 \ifb. 
An uncertainty on the offline reconstructed tau energy scale of $2-4$\%, depending on transverse energy and pseudorapidity, is achieved using two independent methods. The offline tau identification efficiency is measured 
with a precision of 2.5\% for hadronically decaying tau leptons with one associated track, and of 4\% for the case of three associated tracks, inclusive in pseudorapidity and for a visible transverse energy greater than $20$ \GeV.
For hadronic tau lepton decays selected by offline algorithms, the tau trigger identification efficiency is measured with a precision of $2-8$\%, depending on the transverse energy.
The performance of the tau algorithms, both offline and at the trigger level, is found to be stable with respect to the number of concurrent proton-proton interactions 
and has supported a variety of physics results using hadronically decaying tau leptons at ATLAS.
\end{abstract}

\section{Introduction}\label{sec:introduction}
With a mass of 1.777 \GeV\ and a proper decay length of 87 $\mu$m~\cite{PDG}, tau leptons decay either leptonically ($\tau \to \ell\nu_\ell\nu_\tau$, $\ell=e, \mu$) 
or hadronically ($\tau \to \mathrm{hadrons} \; \nu_\tau$, denoted \tauhad) and do so typically before reaching active regions of the ATLAS detector. They can thus only be identified via 
their decay products. In this paper, only hadronic tau lepton decays are considered.
The hadronic tau lepton decays represent 65\% of all possible decay modes~\cite{PDG}. In these, the hadronic decay products are
one or three charged pions in 72\% and 22\% of all cases, respectively.  Charged kaons are present in the majority of the remaining hadronic decays. In 78\% of all hadronic decays, up to one associated neutral pion is also produced.  The neutral and charged hadrons stemming from the tau lepton decay make up the visible decay products of the tau lepton, and are in the following referred to as \tauhadvis.

The main background to hadronic tau lepton decays is from jets of energetic hadrons produced via the 
fragmentation of quarks and gluons. This background is already present at trigger level (also referred to as {\it online} in the following). 
Other important backgrounds are electrons and, to a lesser degree, muons, which can mimic the
signature of tau lepton decays with one charged hadron.
In the context of both the trigger and the offline event reconstruction (shortened to simply {\it offline} in the following), discriminating variables 
 based on the narrow shower shape, the distinct number of charged particle tracks and the displaced tau lepton decay vertex are used.

Final states with hadronically decaying tau leptons are an important part of the ATLAS physics program. Examples are measurements of Standard Model 
processes~\cite{Aad:2012vip,Aad:2012mza,Aad:2012cia,Aad:2011fu,Aad:2011kt}, Higgs boson searches~\cite{Aad:2012mea}, 
searches for new physics such as Higgs bosons in models with extended Higgs sectors~\cite{Aad:2012rjx,Aad:2012tj,Aad:2012cfr}, 
supersymmetry (SUSY)~\cite{Aad:2014mra,Aad:2014yka,Aad:2012ypy}, heavy gauge bosons~\cite{Aad:2012gm} and leptoquarks~\cite{ATLAS:2013oea}. 
This places strong requirements on the \tauhadvis identification algorithms (in the following, referred to as {\it \tauid}): 
robustness and high performance over at least two orders of magnitude in transverse momentum with respect to the beam axis ($\pt$) of 
\tauhadvis, from about 15 \GeV\ (decays of $W$ and $Z$ bosons or scalar tau leptons) to a few hundred \GeV\ (SUSY Higgs boson searches) and up to beyond 1 \TeV\ ($Z'$ searches). At the same time, 
an excellent energy resolution and small energy scale uncertainty are particularly important where resonances decaying to tau leptons need to be 
separated (e.g. $Z \to \tau\tau$ from $H \to \tau\tau$ mass peaks). The triggering for final states which rely exclusively on tau triggers is particularly challenging, e.g. 
$H \to \tau \tau$ where both tau leptons decay hadronically. At the trigger level, in addition to the challenges of offline tau identification, bandwidth and time 
constraints need to be satisfied and the trigger identification is based on an incomplete reconstruction of the event. The ATLAS trigger system, together with the detector and the simulation samples used for the studies presented, are briefly described in Sect.~\ref{sec:detector}. 

The ATLAS offline tau identification uses various discriminating variables combined in Boosted Decision Trees (BDT)~\cite{dt,adaboost} to reject jets and electrons. 
The offline tau energy scale is set by first applying a local hadronic calibration (LC)~\cite{LCCalibration} appropriate for a wide range of objects and then an additional tau-specific correction 
based on simulation. The online tau identification is implemented in three different steps, as is required by the ATLAS trigger system architecture~\cite{cscbook}.
The same identification and energy calibration procedures as for offline are used in the third level of the trigger, while the first and second trigger levels 
rely on coarser identification and energy calibration procedures.
A description of the trigger and offline \tauhadvis reconstruction and identification algorithms is presented in Sect.~\ref{sec:offline}, and the trigger and offline energy calibration algorithms are discussed in Sect.~\ref{sec:tes}.

The efficiency of the identification and the energy scale are measured in dedicated studies using a $Z \to \tau\tau$-enhanced event sample of 
collision data recorded by the ATLAS detector~\cite{atlas_det} at the LHC~\cite{lhc} in 2012 at a centre-of-mass energy of 8 \TeV. This is described in
Sect.~\ref{sec:performance} and Sect.~\ref{sec:tes}. 
Conclusions and outlook are presented in Sect.~\ref{sec:conclusions}. 

\begin{table*}
\begin{center}
\begin{tabular}{llll}
\hline\hline\noalign{\smallskip}
Process & Trigger &  \multicolumn{2}{c}{Requirements at EF [\GeV]} \\
\noalign{\smallskip}\hline\noalign{\smallskip}
$H^{\pm}\rightarrow$\tauhad$\nu$ & \tauhadvis\ + \met            & $\pt(\tau)>29$  & \met $>50$ \\
$H_{\mathrm{SM}}\rightarrow$\tauhad\taulep, $Z\rightarrow$\tauhad\taulep & \tauhadvis\ + $e$        & $\pt(\tau)>20$  & $\pt(e)>18$ \\
                                     & \tauhadvis\ + $\mu$      & $\pt(\tau)>20$  & $\pt(\mu)>15$ \\
$H_{\mathrm{SM}}\rightarrow$\tauhad\tauhad    & \tauhadvis\ + \tauhadvis & $\pt(\tau_1)>29$  & $\pt(\tau_2)>20$ \\
SUSY(\tauhad\tauhad), $H_{\mathrm{SUSY}}\rightarrow$\tauhad\tauhad  & \tauhadvis\ + \tauhadvis & $\pt(\tau_1)>38$  & $\pt(\tau_2)>38$ \\
$Z$'$\rightarrow$\tauhad\tauhad       & \tauhadvis\ + \tauhadvis & $\pt(\tau_1)>100$ & $\pt(\tau_2)>70$ \\
$W$'$\rightarrow$\tauhad$\nu$          & \tauhadvis               & $\pt(\tau)>115$ &  \\
\noalign{\smallskip}\hline\hline
\end{tabular}
\end{center}
\caption{
Tau triggers with their corresponding kinematic requirements. Examples of physics processes targeted by each trigger are also listed, where \tauhad\ and \taulep\ refer to hadronically and leptonically decaying tau leptons, respectively.
}
\label{tab:menu}
\end{table*}

\section{ATLAS detector and simulation}\label{sec:detector}

\subsection{The ATLAS detector}\label{sec:atlas}
The ATLAS detector~\cite{atlas_det} 
consists of an inner tracking system surrounded by a
superconducting solenoid, electromagnetic (EM) and hadronic (HAD) calorimeters, and a muon spectrometer (MS). The inner detector (ID) is immersed in a 2 T axial magnetic field, 
and consists of pixel and silicon microstrip (SCT) detectors inside a transition radiation tracker (TRT), 
providing charged-particle tracking in the region $|\eta|<2.5$.\footnote{
ATLAS uses a right-handed coordinate system with its origin at the nominal interaction point (IP) in the centre of the detector
and the $z$-axis along the beam direction. The $x$-axis points from the IP to the centre of the LHC ring, and the $y$-axis points upward. Cylindrical
coordinates $(r,\phi)$ are used in the transverse $(x,y)$ plane, $\phi$ being the azimuthal angle around the beam direction. The pseudorapidity is defined in
terms of the polar angle $\theta$ as $\eta=-\ln\tan(\theta/2)$.
The distance $\Delta{}R$ in the $\eta$--$\phi$ space is defined as $\Delta{}R=\sqrt{({\Delta\eta})^2 + ({\Delta\phi})^2}$.}
The EM calorimeter uses lead and liquid argon (LAr) as absorber and active material, respectively. In the central rapidity region, the EM calorimeter is divided in three layers, one of them segmented in thin $\eta$ strips for optimal $\gamma / \pi^0$ separation, and completed by a presampler layer for $|\eta| < 1.8$. Hadron calorimetry is based on different detector technologies, with scintillator tiles ($|\eta| < 1.7$) or LAr ($1.5 < |\eta| < 4.9$) as active medium, and with steel, copper, 
or tungsten as the absorber material. The calorimeters provide coverage within $|\eta|<4.9$.
The MS consists of superconducting air-core toroids, a system of trigger chambers covering the range $|\eta|<2.4$, and high-precision tracking chambers allowing muon momentum measurements 
within $|\eta|<2.7$.  

Physics objects are identified using their specific detector signatures; electrons are reconstructed by matching a track from the ID to an energy deposit in the calorimeters~\cite{egammaPaper1,egammaPaper2},
while muons are reconstructed using tracks from the MS and ID~\cite{muonPaper}.  Jets are reconstructed using the anti-$k_{t}$ algorithm~\cite{antikt} with a distance parameter $R = 0.4$.
Three-dimensional clusters of calorimeter cells called {\it TopoClusters}~\cite{TopoClusters}, calibrated using a local hadronic calibration~\cite{LCCalibration},
serve as inputs to the jet algorithm.  The missing transverse momentum (with magnitude \met) is computed from the combination of all reconstructed physics objects and the
remaining calorimeter energy deposits not included in these objects ~\cite{met}.

The ATLAS trigger system \cite{cscbook} 
consists of three levels; the first level (L1) is hardware-based while the second (L2) and third (Event Filter, EF) levels 
are software-based. The combination of L2 and the EF are referred to as the high-level trigger (HLT). The L1 
trigger identifies regions-of-interest (RoI) using information from the calorimeters and the muon spectrometer. The delay between a beam crossing 
and the trigger decision (latency) is approximately 2~$\mu$s at L1. The L2 system typically takes the RoIs produced by L1 as input and refines the 
quantities used for selection after taking into account the information from all subsystems. The latency at L2 is on average 40~ms, but can
be as large as 100~ms at the highest instantaneous luminosities.
At the EF level, algorithms similar to those run in the offline reconstruction are used to select interesting events with an average latency 
of about 1~s. 

During 2012, the ATLAS detector was operated with a data-taking efficiency greater than $95 \%$. The highest peak luminosity obtained was 
$8 \cdot 10^{33}~ \mathrm{cm} ^{-2} \mathrm{s}^{-1}$ at the end of 
2012. 
The observed average number of pile-up interactions (meaning generally soft proton--proton interactions, superimposed on one hard proton--proton interaction) per
bunch crossing in 2012 was 20.7. At the end of the data-taking period, the trigger system was routinely working with an average (peak) output rate of 700 Hz (1000 Hz).

\subsection{Tau trigger operation}\label{sec:triggerop}
In 2012, a diverse set of tau triggers was implemented, using requirements on different final state configurations
to maximize the sensitivity to a large range of physics processes. These triggers are listed in Table~\ref{tab:menu}, along
with the targeted physics processes and the associated kinematic requirements on the triggered objects.
For the double hadronic triggers, in the lowest threshold version (29 and 20 GeV requirement on transverse momentum for the two \tauhadvis)
 two main criteria are applied: isolation at L1\footnote{A detailed definition of the isolation requirement is provided in Sect.~\ref{sec:triggerreco}.}, and full tau identification at the HLT. 
The isolation requirement is dropped for the intermediate threshold version, and both criteria are dropped in favour of a looser (more than 95\% efficient), non-isolated trigger for the version with the highest thresholds.

As the typical rejection rates of \tauhadvis identification algorithms against the dominant multi-jet backgrounds are considerably smaller than those of
electron or muon identification algorithms, \tauhadvis triggers must have considerably higher \pt\ requirements
in order to maintain manageable trigger rates. Therefore, most analyses using low-\pt\ \tauhadvis in 2012 depend on the
use of triggers which identify other objects. However, by combining tau trigger requirements with
requirements on other objects, lower thresholds can be accommodated for the tau trigger objects as well as the other objects.

\begin{figure}[htbp]
\begin{center}
\subfigure[]{
\includegraphics[width=0.47\textwidth]{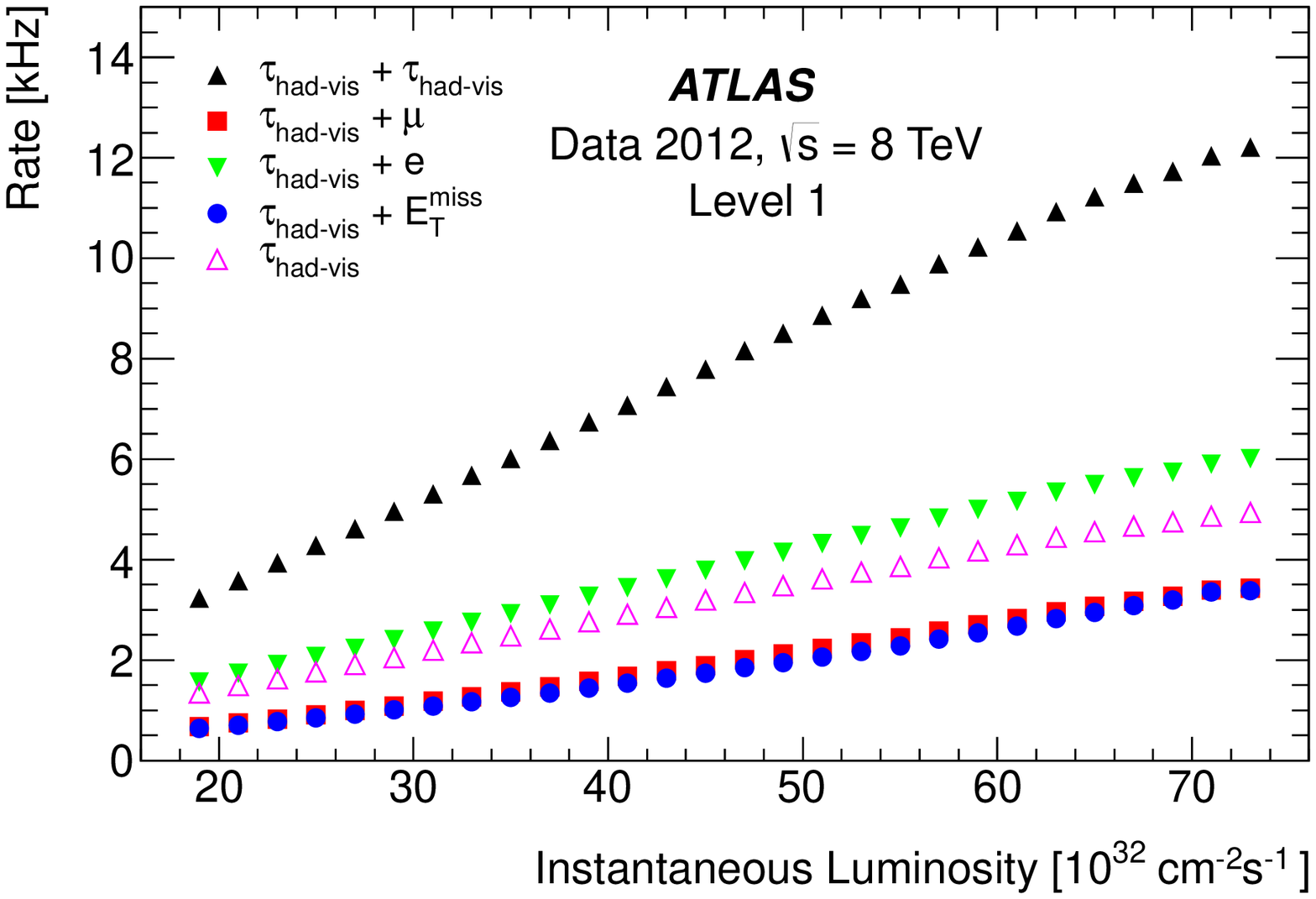}
}
\subfigure[]{
\includegraphics[width=0.47\textwidth]{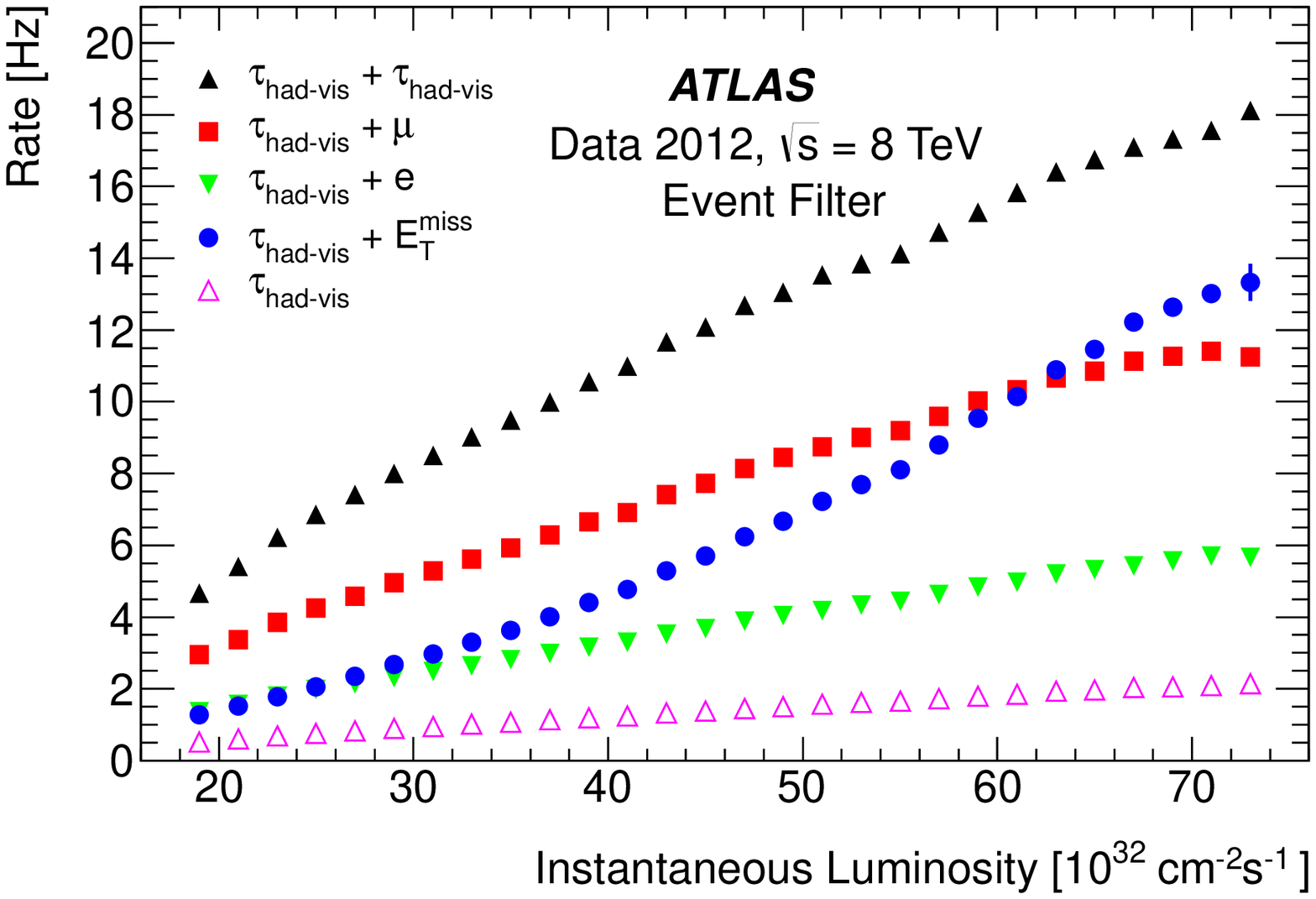}
}
\caption{
Tau trigger rates at (a) Level 1 and (b) Event Filter as a function of the instantaneous luminosity for $\sqrt{s}=8$\,\TeV. 
The triggers shown are described in Table~\ref{tab:menu}, with the \tauhadvis +\tauhadvis being the rate for the lowest threshold trigger reported in the table.
The rates for the higher threshold triggers are approximately three and five times lower at L1 and HLT, respectively, and are partially
included in the rate of the lowest threshold item.
\label{fig:trigrate2012}}
\end{center}
\end{figure}

Figure~\ref{fig:trigrate2012} shows the tau trigger rates at L1 and the EF as a function of the instantaneous luminosity during the 8 \TeV\ LHC operation. 
The trigger rates do not increase more than linearly with the luminosity, due the robust performance
of the trigger algorithms under different pile-up conditions.
The only exception is the \tauhadvis\ + \met\ trigger,
where the extra pile-up associated with the higher luminosity leads to a degradation of the resolution of the reconstructed
event \met.  At the highest instantaneous luminosities, the rates are affected by deadtime in the readout systems,
leading to a general drop in the rates.


\subsection{Simulation and event samples}\label{sec:samples}
The optimization and measurement of tau performance requires simulated events. 
Events with $Z/\gamma^*$ and $W$ boson production 
were generated using {\sc alpgen}~\cite{Alpgen} interfaced to {\sc herwig}~\cite{Herwig} or {\sc Pythia6}~\cite{Pythia6} 
for fragmentation, hadronization and underlying-event (UE) modelling.
In addition, $Z \to \tau \tau$ and $W \to \tau\nu$ events were generated using {\sc Pythia8}~\cite{Pythia8}, and provide a larger statistical sample for the studies.
For optimization at high \pt{}, $Z' \to \tau \tau$ with $Z'$ masses between 250 \GeV\ and 1250 \GeV\ were generated with {\sc Pythia8}.
 Top-quark-pair as well as single-top-quark events were generated with {\sc mc@nlo}+{\sc herwig}~\cite{MCNLO}, with the exception of 
t-channel single-top production, where {\sc AcerMC}+{\sc Pythia6}~\cite{AcerMC} was used. $WZ$ and $ZZ$ diboson events were generated with {\sc herwig}, and $WW$ events with {\sc alpgen}+{\sc herwig}. 
In all samples with $\tau$ leptons, except for those simulated with {\sc Pythia8},
{\sc Tauola}~\cite{Tauola} was used to model the $\tau$ decays, and
{\sc Photos}~\cite{Photos} was used for soft QED radiative corrections to particle decays.

All events were produced using CTEQ6L1 ~\cite{ct6} parton distribution functions (PDFs) 
except for the {\sc mc@nlo} events, which used CT10 PDFs~\cite{ct10}.
The UE simulation was tuned using collision data. {\sc Pythia8} events employed the AU2 tune~\cite{PUB-2012-003}, 
{\sc herwig} events the AUET2 tune~\cite{PUB-2011-008}, while {\sc alpgen+Pythia6} used the Perugia2011C tune~\cite{p2011tune} and 
{\sc AcerMC+Pythia6} the AUET2B tune~\cite{PUB-2011-009}. 

The response of the ATLAS detector was simulated using GEANT4~\cite{Atlassim,geant} with the hadronic-shower model QGSP\_BERT~\cite{Folger:2003sb,Bertini} as baseline. 
Alternative models (FTFP\_BERT~\cite{FTF} and QGSP) were used to estimate systematic uncertainties. 
Simulated events were overlaid
with additional minimum-bias events generated with {\sc Pythia8} to account for the effect of multiple interactions 
occurring in the same and neighbouring bunch crossings
(called pile-up). Prior to any analysis, 
the simulated events were reweighted such that the distribution of the number of pile-up interactions matched that in data.
The simulated events were reconstructed with 
the same algorithm chain as used for collision data.

\section{Reconstruction and identification of hadronic tau lepton decays}\label{sec:offline}

In the following, the \tauhadvis reconstruction and identification at online and offline level are described. 
The trigger algorithms were optimized with respect to hadronic tau decays identified by the offline algorithms. This typically leads to online algorithms resembling their offline counterparts as closely as possible with the information available at a given 
trigger level. To reflect this, the details of the offline reconstruction and identification are described first, and then a discussion of the trigger algorithms follows, 
highlighting the differences between the two implementations.

\subsection{Reconstruction}\label{sec:recoid}
The \tauhadvis reconstruction algorithm is seeded by calorimeter energy deposits which have been reconstructed as individual jets.  Such jets
are formed using the anti-$k_t$ algorithm with a distance parameter of $R=0.4$, using calorimeter TopoClusters as inputs. To seed a \tauhadvis candidate,
a jet must fulfil the requirements of $\pt>10$ \GeV\ and $\abseta<2.5$. 
Events must have a reconstructed primary vertex with at least three associated tracks.  In events with multiple primary vertex candidates, the primary vertex is chosen to be the one with the highest $\Sigma p^2_\mathrm{T, tracks}$ value.
In events with multiple simultaneous interactions, the chosen primary vertex does not always correspond to the vertex at which the tau lepton is produced.
To reduce the effects of pile-up and increase reconstruction efficiency, the tau lepton production vertex is identified, amongst the previously reconstructed primary vertex candidates in the event.

The tau vertex (TV) association algorithm uses as input all tau candidate tracks which have $\pt>1$ \GeV{}, satisfy quality criteria based on the number of hits
in the ID, and are in the region $\Delta R<0.2$ around the jet seed direction; no impact parameter requirements are applied.
The $p_{T}$ of these tracks is summed and the primary vertex candidate to which the largest fraction of the $p_{T}$ sum is matched to is chosen as the TV~\cite{JVF}.

This vertex is used in the following to determine the \tauhadvis direction, to associate tracks and to build the 
coordinate system in which identification variables are calculated. In $Z \to \tau\tau$ events, the TV coincides with the highest $\Sigma p^2_\mathrm{T,tracks}$ vertex (for the pile-up profile observed during 2012) roughly 90\% of the time. For physics analyses which require higher-\pt{} objects, the two coincide in more than 99\% of all cases.

The \tauhadvis three-momentum is calculated by first computing $\eta$ and $\phi$ of the barycentre of the TopoClusters of the jet seed, calibrated at the LC scale, 
assuming a mass of zero for each constituent. The four-momenta of all clusters in the region $\Delta R<0.2$ around the barycentre are recalculated using 
the TV coordinate system and summed, resulting in the momentum magnitude $p^\mathrm{LC}$ and a \tauhadvis direction. The \tauhadvis mass is defined to be zero.

Tracks are associated with the \tauhadvis if they are in the {\it core region} $\Delta R<0.2$ around the \tauhadvis 
direction and satisfy the following criteria: $\pt>1$ \GeV, at least two 
associated hits in the pixel layers of the inner detector, and at least seven hits in total in the pixel and the SCT layers. Furthermore, 
requirements are imposed on the distance of closest approach of the track to the TV in the transverse plane, $|d_0|<1.0$ mm, and longitudinally, $|z_0 \sin \theta|<1.5$ mm. 
When classifying a \tauhadvis candidate as a function of its number of associated tracks, the selection listed above is used.
Tracks in the {\it isolation region} $0.2<\Delta R<0.4$ are used for the calculation of 
identification variables and are required to satisfy the same selection criteria.

A $\pi^0$ reconstruction algorithm was also developed. In a first step, the algorithm measures the number of reconstructed neutral pions (zero, one or two), $N_{\mathrm{\pi^0}}$, in the core region, by looking at global tau features measured using strip layer and calorimeter quantities, and track momenta, combined in BDT algorithms.
In a second step, the algorithm combines the kinematic information of tracks and of clusters likely stemming from $\pi^0$ decays. A candidate $\pi^0$ decay is composed of up to two clusters among those found in the core region of \tauhadvis candidates. Cluster properties are used to assign a $\pi^0$ likeness score to each cluster found in the core region, after subtraction of the contributions from pile-up, the underlying event and electronic noise (estimated in the isolation region). 
Only those clusters with the highest scores are used, together with the reconstructed tracks in the core region of the \tauhadvis candidate, to define the input variables for tau identification described in the next section.

\subsection{Discrimination against jets}
\label{offlineid}

The reconstruction of \tauhadvis candidates provides very little rejection against the jet background. 
Jets in which the dominant 
particle\footnote{This is often interpreted as the parton initiating the jet or the highest-\pt{} parton within a jet; 
however, none of these concepts can be defined unambiguously.}
is a quark or a gluon are referred to as {\it quark-like} and {\it gluon-like} jets, respectively. Quark-like jets 
are on average more collimated and have fewer tracks and thus the discrimination from \tauhadvis is less effective than for gluon-like jets.
Rejection against jets 
is provided in a separate identification step using discriminating variables based on the tracks and TopoClusters (and cells linked to them) found in the core 
or isolation 
region around the \tauhadvis candidate direction. 
The calorimeter measurements provide information about the longitudinal and lateral shower shape and the $\pi^0$ content of tau hadronic decays.

The full list of discriminating variables used for tau identification is given below and is summarized in Table~\ref{tab:variables_used}.
 
\begin{description}

\item[{\bf Central energy fraction (\coreEnergyFrac{}):}] Fraction of transverse energy deposited in the region
  $\Delta R < 0.1$ with respect to all energy deposited in the region $\Delta R < 0.2$ around the \tauhadvis candidate 
calculated by summing the energy deposited in all cells belonging to TopoClusters with 
a barycentre in this region, calibrated at the EM energy scale. 
Biases due to pile-up contributions are removed using a correction based on the number of reconstructed primary vertices in the event.

\item[{\bf Leading track momentum fraction (\leadTrackMomFrac{}):}]
The transverse momentum of the highest-\pt{} charged particle in the core region of
the \tauhadvis candidate, divided by the transverse energy sum, calibrated at the EM energy scale, deposited in all cells belonging to 
TopoClusters in the core region. A correction depending on the number of reconstructed primary vertices in the event is applied to this fraction, making the resulting variable pile-up independent.

\item[{\bf Track radius (\trackRadius{}):}] \pt-weighted distance of the associated tracks to the \tauhadvis direction,
using all tracks in the core and isolation regions.

\item[{\bf Leading track IP significance (\ipSigLeadTrk{}):}] Transverse impact parameter of
 the highest-\pt{} track in the core region, calculated with respect to the TV, divided by its estimated uncertainty. 

\item[{\bf Number of tracks in the isolation region (\numIsoTrack{}):}] Number of tracks associated with the \tauhadvis in the region $0.2<\Delta R<0.4$.

\item[{\bf Maximum $\Delta R$ (\dRmax{}):}] The maximum $\Delta R$ between a track associated
 with the \tauhadvis candidate and the \tauhadvis direction. Only tracks in the core region are considered.
           
\item[{\bf Transverse flight path significance (\transFlightSig{}):}] The decay length of the secondary vertex (vertex reconstructed from the tracks associated with the core region of the \tauhadvis candidate) in the transverse plane, calculated with respect to the TV, divided by its estimated uncertainty. It is defined only for multi-track \tauhadvis candidates.

\item[{\bf Track mass (\trackMass{}):}] Invariant mass calculated from the sum of the four-momentum of all tracks in the core and isolation regions,
assuming a pion mass for each track.

\item[{\bf Track-plus-$\pi^0$-system mass (\trackPizeroMass{}):}] Invariant mass of the system composed of the tracks and $\pi^0$ mesons in the core region.

\item[{\bf Number of $\pi^0$ mesons ($N_{\mathrm{\pi^0} } $):}] Number of $\pi^0$ mesons reconstructed in the core region.

\item[{\bf Ratio of track-plus-$\pi^0$-system \pt{} (\Etratio{}):}] Ratio of the \pt{} estimated using the track + $\pi^0$ information to the calorimeter-only measurement.

\end{description}

\begin{table}[htp]
\begin{center}
\begin{tabular}{ccccc}
\hline\hline\noalign{\smallskip}
Variable & \multicolumn{2}{c}{Offline} & \multicolumn{2}{c}{Trigger} \\
         &  1-track & 3-track & 1-track & 3-track \\
\noalign{\smallskip}\hline\noalign{\smallskip}
\coreEnergyFrac{}    & $\bullet$ & $\bullet$ & $\bullet$ & $\bullet$ \\
\leadTrackMomFrac{}  & $\bullet$ & $\bullet$ & $\bullet$ & $\bullet$ \\
\trkAvgDist{}        & $\bullet$ & $\bullet$ & $\bullet$ & $\bullet$ \\
\ipSigLeadTrk{}      & $\bullet$ &           & $\bullet$ &           \\
\numIsoTrack{}       & $\bullet$ &           & $\bullet$ &           \\
\dRmax{}             &           & $\bullet$ &           & $\bullet$ \\
\trkFlightPathSig{}  &           & $\bullet$ &           & $\bullet$ \\
\massTrkSys{}        &           & $\bullet$ &           & $\bullet$ \\
\massTrkPizeroSys{}  & $\bullet$ & $\bullet$ &           &           \\
$N_{\mathrm{\pi^0}}$ & $\bullet$ & $\bullet$ &           &           \\
\Etratio{}           & $\bullet$ & $\bullet$ &           &           \\
\hline\hline\noalign{\smallskip}
\end{tabular}
\end{center}
\caption{
Discriminating variables used as input to the \tauid algorithm 
at offline reconstruction and at trigger level, for 1-track and 3-track candidates.
The bullets indicate whether a particular variable is used for a given selection.
The $\pi^{0}$-reconstruction-based variables, \massTrkPizeroSys{}, $N_{\mathrm{\pi^0}}$, \Etratio{} are not used in the trigger.
}
\label{tab:variables_used}
\end{table}

\begin{figure}[!htbp]
\begin{center}
\subfigure[]{
        \includegraphics[width=0.47\textwidth]{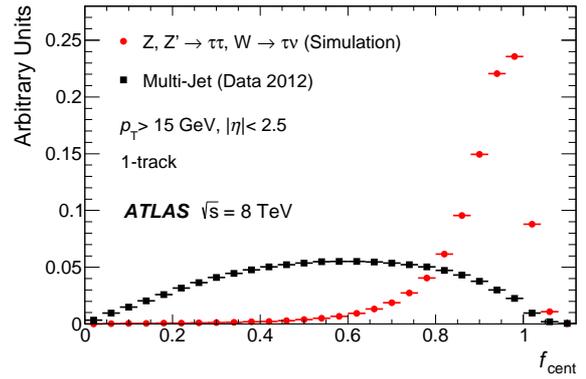}
}
\subfigure[]{
        \includegraphics[width=0.47\textwidth]{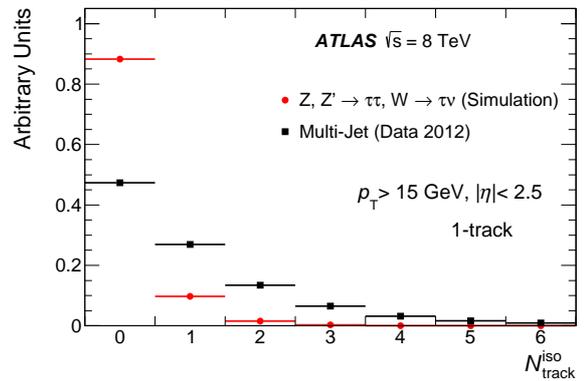}
}
\caption{
Signal and background distribution for the 1-track \tauhadvis decay offline tau identification variables (a) \coreEnergyFrac{} and (b) \numIsoTrack{}.
For signal distributions, 1-track \tauhadvis decays are matched to true generator-level \tauhadvis in simulated events, while
the multi-jet events are obtained from the data.
}\label{fig:tauidvariables1P}
\end{center}
\end{figure}

\begin{figure}[!htbp]
\begin{center}
\subfigure[]{
        \includegraphics[width=0.47\textwidth]{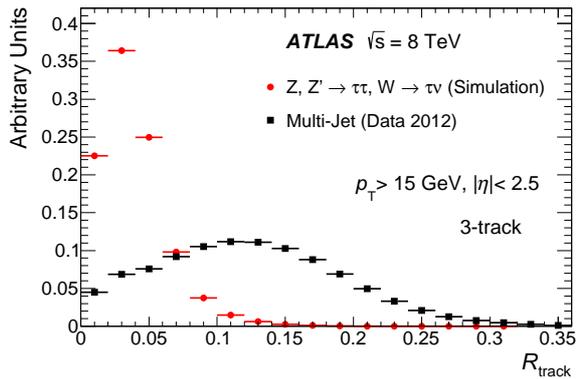}
}
\subfigure[]{
        \includegraphics[width=0.47\textwidth]{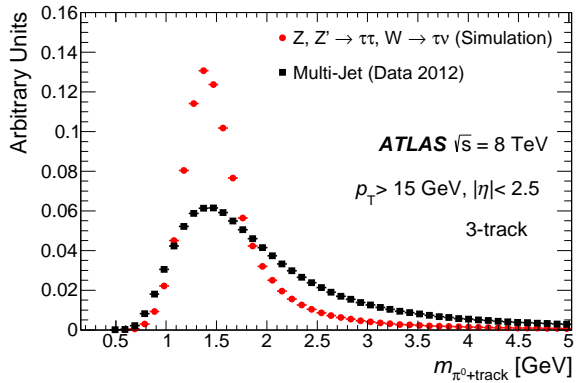}
}
\caption{
Signal and background distribution for the 3-track \tauhadvis decay offline tau identification variables (a) \trackRadius{} and (b) \massTrkPizeroSys{}.
For signal distributions, 3-track \tauhadvis decays are matched to true generator-level \tauhadvis in simulated events, while
the multi-jet events are obtained from data.
}\label{fig:tauidvariables3P}
\end{center}
\end{figure}

The distributions of some of the important discriminating variables listed in Table~\ref{tab:variables_used}
are shown in Figs.~\ref{fig:tauidvariables1P} and ~\ref{fig:tauidvariables3P}. 

Separate BDT algorithms are trained for 1-track and 3-track \tauhadvis decays using a combination of simulated tau leptons in $Z$, $W$ and $Z'$ decays. 
For the jet background, large collision data samples collected by jet triggers, referred from now on as the multi-jet data samples, are used. 
For the signal, only reconstructed \tauhadvis candidates matched to the true (i.e., generator-level) visible hadronic tau decay 
products in the region around $\Delta R < 0.2$ with $p^{\textrm{true}}_{\textrm{T,vis}} > 10$\,\GeV{} and $|\eta^{\textrm{true}}_{\textrm{vis}}| < 2.3$ are used.
In the following, the signal efficiency is defined as the fraction of true visible hadronic tau decays with $n$ 
charged decay products, which are reconstructed with $n$ associated tracks and satisfy \tauid criteria. The background efficiency is the fraction of reconstructed \tauhadvis candidates with $n$ associated tracks which satisfy \tauid criteria, measured in a background-dominated sample. 

\begin{figure}[htbp]
\begin{center}
\subfigure[]{
        \includegraphics[width=0.47\textwidth]{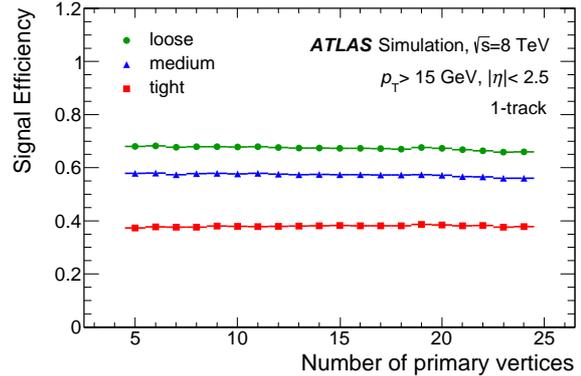}
}
\subfigure[]{
        \includegraphics[width=0.47\textwidth]{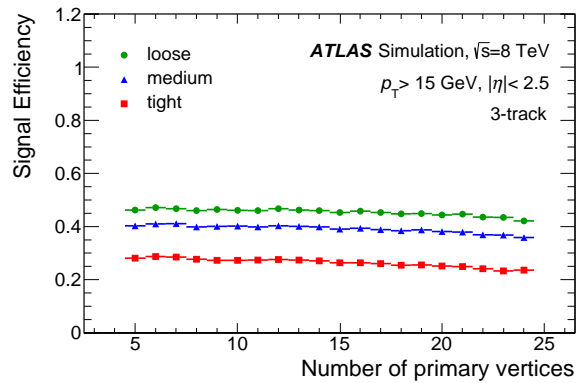}
}
\caption{
Offline tau identification efficiency dependence on the number of reconstructed interaction vertices, for (a) 1-track and (b) 3-track \tauhadvis decays matched to 
true \tauhadvis (with corresponding number of charged decay products) from SM and exotic processes in simulated data. Three working points, corresponding to different \tauid efficiency values, are shown. 
}\label{fig:sigeffvsnvtx}
\end{center}
\end{figure}

\begin{figure}[htbp]
\begin{center}
\subfigure[]{
        \includegraphics[width=0.47\textwidth]{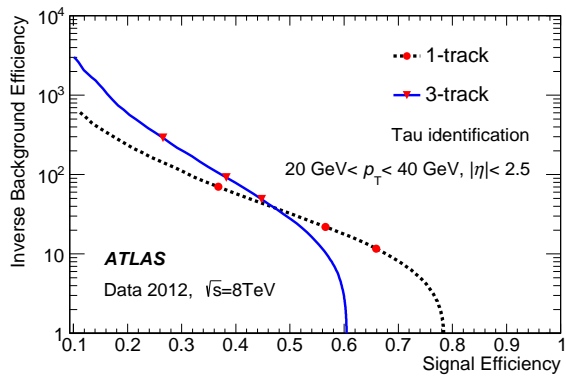}
}
\subfigure[]{
        \includegraphics[width=0.47\textwidth]{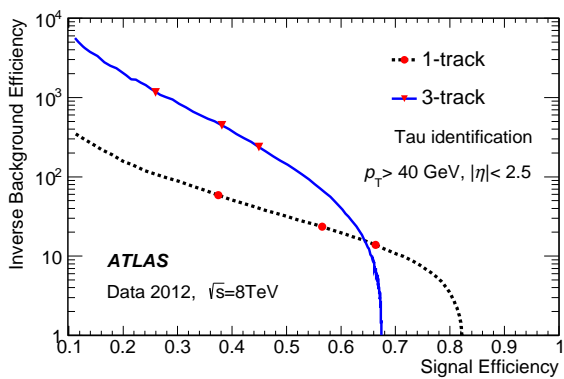}
}
\caption{
Inverse background efficiency versus signal efficiency for the offline \tauid, for (a) a low-\pt\ and (b) a high-\pt\ \tauhadvis range. Simulation samples for signal include a mixture of $Z$, $W$ and $Z'$ production processes, while data from multi-jet events is used for background. The red markers correspond to the three working points mentioned in the text. The signal efficiency shown corresponds to the total efficiency of \tauhadvis decays to be reconstructed as 1-track or 3-track and pass tau identification selection.}\label{fig:roc}
\end{center}
\end{figure}

Three working points, labelled {\it tight}, {\it medium} and {\it loose}, are provided, corresponding to different \tauid efficiency values. Their signal efficiency values (defined with respect to 1-track or 3-track reconstructed \tauhadvis candidates matched to true \tauhadvis) can be seen in Fig.~\ref{fig:sigeffvsnvtx}. The requirements on the BDT score 
are chosen such that the resulting efficiency is independent of the true \tauhadvis \pt. Due to the choice of input variables, the \tauid also shows 
stability with respect to the pile-up conditions as shown in Fig.~\ref{fig:sigeffvsnvtx}.
The performance of the \tauid algorithm in terms of the inverse background efficiency versus the signal efficiency is shown in Fig.~\ref{fig:roc}. At low transverse momentum of the \tauhadvis candidates, 40\% signal efficiency for an inverse background efficiency of 60 is achieved. The signal efficiency saturation point, visible in these curves, stems from the reconstruction efficiency for a true \tauhadvis with one or three charged decay products to be reconstructed as a 1-track or 3-track \tauhadvis candidate. The main sources of inefficiency are track reconstruction efficiency due to hadronic interactions and migration of the number of reconstructed tracks due to conversions or underlying-event tracks being erroneously associated with the tau candidate.
 
\subsection{Tau trigger implementation}\label{sec:triggerreco}
The tau reconstruction at the trigger level has differences with respect to its offline
counterpart due to the technical limitations of the trigger system.  At L1, no inner detector track
reconstruction is available, and the full calorimeter granularity cannot be accessed.
Latency limits at L2 prevent the use of the TopoCluster algorithm, 
and only allow the candidate reconstruction to be performed within the given
RoI.  At the EF, the same tau reconstruction and identification methods as offline
are used, except for the $\pi^{0}$ reconstruction. In this section, the details of the
tau trigger reconstruction algorithm are described.

\paragraph{Level 1}
At L1, the \tauhadvis candidates are selected using calorimeter energy deposits. Two calorimeter regions are defined by the tau trigger
for each candidate, using trigger towers in both the EM and HAD calorimeters: the core region, and an isolation region around this core.
The trigger towers have a granularity of $\Delta\eta\times\Delta\phi=0.1\times0.1$ with a 
coverage of $\abseta < 2.5$. The core region is defined as a square of $2\times2$ trigger towers, corresponding to $0.2\times0.2$ in $\Delta\eta\times\Delta\phi$ space.
The \et\ of a \tauhadvis candidate at L1 is taken as the sum of the transverse energy in the two most energetic neighbouring central towers in the EM calorimeter core region,
and in the $2\times2$ towers in the HAD calorimeter, all calibrated at the EM scale. For each \tauhadvis candidate, the EM isolation is calculated as the transverse energy deposited in the annulus between $0.2\times0.2$ and $0.4\times0.4$ in 
the EM calorimeter.

To suppress background events and thus reduce trigger rates, an EM isolation energy of less than 4 \GeV\ is required for the lowest \et\ threshold
at L1. Hardware limitations prevent the use of an \et-dependent selection.
This requirement reduces the efficiency of \tauhadvis events by less than 2\% over most of the kinematic range. Larger efficiency losses occur for \tauhadvis events at high
\et\ values; those are recovered through the use of triggers with higher \et\ thresholds but without any isolation requirements.

The energy resolution at L1 is significantly lower than at the offline level. This is due to the fact that all cells in a trigger tower 
are combined without the use of sophisticated clustering algorithms and without \tauhadvis-specific energy calibrations. Also, the coarse energy and geometrical
position granularity limits the precision of the measurement. These effects lead to a significant signal efficiency loss for low-\et\ \tauhadvis candidates. 

\paragraph{Level 2}
At L2, \tauhadvis candidate RoIs from L1 are used as seeds to reconstruct both the calorimeter- and tracking-based
observables associated with each \tauhadvis candidate. The events are then selected based on an identification algorithm
that uses these observables. The calorimeter observables associated with the \tauhadvis candidates are calculated using calorimeter
cells, where the electronic and pile-up noise are subtracted in the energy calibration.
The centre of the \tauhadvis energy deposit is taken as the 
energy-weighted sum of the cells collected in the region $\Delta R<0.4$ around the L1 seed. The transverse energy of the \tauhadvis is
calculated using only the cells in the region $\Delta R<0.2$ around its centre.

To calculate the tracking-based observables, a fast tracking algorithm~\cite{Aad:2012xs} is applied, using only hits from the pixel and
SCT tracking layers. Only tracks satisfying $\pt > 1.5$ \GeV\ and located in the region $\Delta R<0.3$ around the L2 calorimeter \tauhadvis direction are used. 
The tracking efficiency with respect to offline  
reaches a plateau of 99\% at 2 \GeV\ (with an efficiency of about 98\% at 1.5 \GeV). 
The fast tracking algorithm required an average of 37~ms to run at the highest pile-up conditions at peak 
luminosity in 2012 (approximately forty pile-up interactions). 

\begin{figure}[htbp]
\begin{center}
\includegraphics[width=0.47\textwidth]{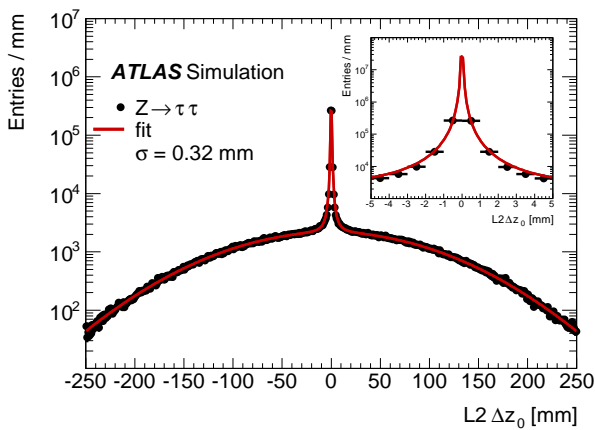}
\caption{
Distribution of $\Delta z_{0}$ for the tau trigger at L2 in simulated \Ztau\ events with an average of eight interactions per bunch crossing. The wide 
Gaussian distribution corresponds to pile-up tracks while the central peak, displayed in the upper-right corner, corresponds to 
the main interaction tracks. A Breit--Wigner function is fitted to the central peak and 68\% of
the signal events are found within a distance $\sigma$ = 0.32 mm from the peak.}
\label{fig:deltaz0cut}
\end{center}
\end{figure}

As there is no vertex information available at this stage, an alternative approach is used to reject tracks coming from
pile-up interactions. A requirement is placed on the $\Delta z_{0}$ between a candidate track and the highest-\pt\ track inside the RoI.
The distribution of $\Delta z_{0}$ is shown in Fig.~\ref{fig:deltaz0cut} for simulated \Ztau\ events with an average of eight interactions per bunch crossing. 
High values of $\Delta z_{0}$ typically correspond to pile-up tracks while the central peak corresponds to the main interaction tracks.

The $\Delta z_{0}$ distribution is fit to the sum of a Breit--Wigner function to describe the central peak and a Gaussian function to describe the broad distribution from tracks in pile-up events.
The half-width of the Breit--Wigner $\sigma$=0.32~mm is taken as the point where 68\% of the signal events are included in the central peak.
A dependence of the trigger variables on pile-up conditions is minimized by considering only tracks within 
$-2$~mm $<\Delta z_{0}<$ 2~mm and $\Delta R<0.1$ with respect to the highest-\pt\ track. 

Track isolation requirements are applied to \tauhadvis candidates to increase background rejection.
For multi-track candidates (candidates with two or three associated tracks, defined to be as inclusive
as possible with respect to their offline counterpart), the ratio of the sum of the track \pt\ in $0.1<\Delta R<0.3$
to the sum of the track \pt\ in $\Delta R<0.1$ is required to be lower than 0.1.  Any 1-track candidate with
a reconstructed track in the isolation region is rejected.

In the last step, identification variables combining calorimeter and track information are built as described in Sect.~\ref{offlineid}. 
The calorimeter-based isolation variable \coreEnergyFrac{} uses an expanded cone size of $\Delta R < 0.4$ without the pile-up correction
term to estimate the fraction of transverse energy deposited in the region $\Delta R < 0.1$ around the \tauhadvis candidate. 
The variables \leadTrackMomFrac{} and \trackRadius{}, measuring respectively the ratio of the transverse momentum of the leading \pt{} track
to the total transverse energy (calibrated at the EM energy scale)  and the \pt-weighted distance of the associated tracks to the \tauhadvis direction, are calculated
using selected tracks in the region $\Delta R < 0.3$ around the highest-\pt\ track. 
Cuts on the chosen identification variables are optimized to provide an inverse background efficiency of roughly ten while keeping the signal efficiency as high as possible
(approximately 90\% with respect to the offline {\it medium} \tauid).

\paragraph{Event Filter}
At the EF level, the \tauhadvis reconstruction is very similar to the offline version. First, the TopoCluster reconstruction and calibration algorithms are
run within the RoI. Then, track reconstruction inside the RoI is performed using the EF tracking algorithm. In the last step, the full offline \tauhadvis reconstruction algorithm is used. 
The EF tracking is almost 100\% efficient over the entire \pt\ range with respect to the offline reconstructed tracks. It is, however, considerably slower than
the L2 fast tracking algorithm, requiring about 200~ms per RoI under severe pile-up conditions (forty pile-up interactions). The TopoClustering algorithms need only about 15~ms. 

The \tauhadvis\ candidate four-momentum and input variables to the EF tau identification are then calculated. The main difference with respect to the offline tau 
reconstruction is that $\pi^{0}$-reconstruction-based input variables (\massTrkPizeroSys{}, $N_{\mathrm{\pi^0}}$ and \Etratio{}) are not used; the methodology
to compute these variables had not yet been developed when
the trigger was implemented. Furthermore, no pile-up correction is applied to the input variables at trigger level.

Since full-event vertex reconstruction is not available at trigger level (vertices are only formed using the tracks in a given RoI), the selection requirements applied to the input tracks are also
different with respect to the offline \tauhadvis reconstruction. Similarly to L2, the $\Delta z_{0}$ requirement for tracks is computed with respect to the leading track, 
and loosened to 1.5 mm with respect to the offline requirement. The $\Delta d_{0}$ requirement is calculated with respect to the vertex found inside of the RoI,
and is loosened to 2 mm.

\begin{figure}[htbp]
\begin{center}
\includegraphics[width=0.47\textwidth]{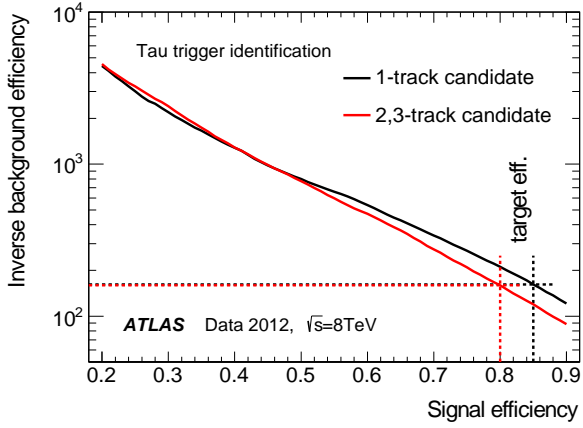}
\caption{
Inverse background efficiency versus signal efficiency for the tau trigger at the EF level, for \tauhadvis candidates
which have satisfied the L1 requirements. The signal efficiency is defined with respect to offline {\it medium}
\tauid\ \tauhadvis candidates matched at generator level, and the inverse background efficiency is calculated in a multi-jet sample. 
}
\label{fig:trigmva}
\end{center}
\end{figure}

A BDT with the input variables listed in 
Table~\ref{tab:variables_used} is used to suppress the backgrounds from jets misidentified as \tauhadvis. 
The BDT was trained on 1- and 3-track \tauhadvis candidates with simulated $Z$, $W$ and $Z'$ events for the signal and data multi-jet samples for the background, respectively.
Only events passing an L1 tau trigger matched with an offline reconstructed \tauhadvis with $\pt>15$\,\GeV\ and 
$\abseta<2.2$ are used, where the {\it medium} identification is required for the \tauhadvis candidates. 
For the signal, in addition, a geometrical matching to a true \tauhadvis is required. 
The performance of the EF tau trigger is presented in Fig.~\ref{fig:trigmva}. 
The signal efficiency is defined with respect to offline reconstructed \tauhadvis candidates matched at generator level, 
and the inverse background efficiency is calculated in a multi-jet sample. 
The working points are chosen to obtain a signal efficiency of 85\% and 80\% with respect to the offline {\it medium}
candidates for 1-track and multi-track candidates respectively, where the inverse background efficiency is of the order
of 200 for the multi-jet sample.

\subsection{Discrimination against electrons and muons}

Additional dedicated algorithms are used to discriminate \tauhadvis from electrons and muons.  These algorithms are only used offline.

\begin{figure}[!htbp]
\begin{center}
\subfigure[]{
  \includegraphics[width=0.47\textwidth]{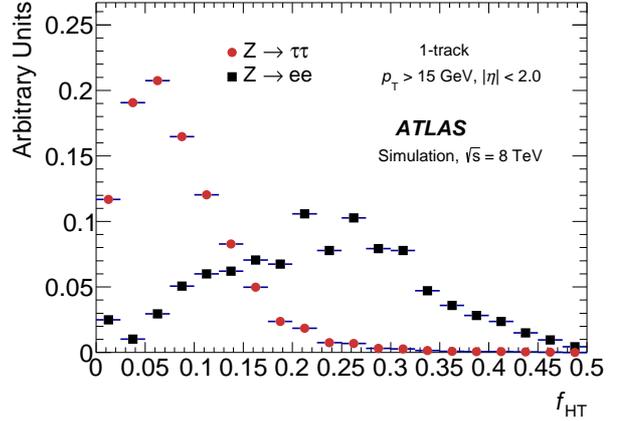}
}
\subfigure[]{
  \includegraphics[width=0.47\textwidth]{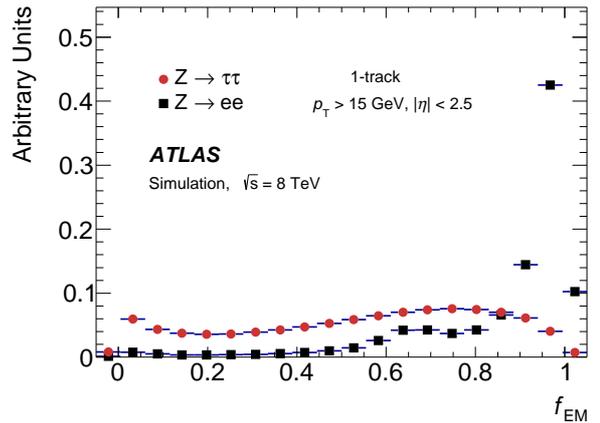}
}
\caption{
Signal and background distribution for two of the electron veto variables, (a) $f_{\mathrm{HT}}$ and (b) $f_{\mathrm{EM}}$. Candidate 1-track \tauhadvis decays are required to not overlap with a reconstructed electron candidate which passes tight electron identification~\cite{egammaPaper2}. 
For signal distributions, 1-track \tauhadvis decays are matched to true generator-level \tauhadvis in simulated $Z\rightarrow\tau\tau$ events, while
the electron contribution is obtained from simulated $Z\rightarrow ee$ events where 1-track \tauhadvis decays are matched to true generator-level electrons.
}\label{fig:evetovars}
\end{center}
\end{figure}

\begin{figure}[htbp]
\begin{center}
 \includegraphics[width=0.47\textwidth]{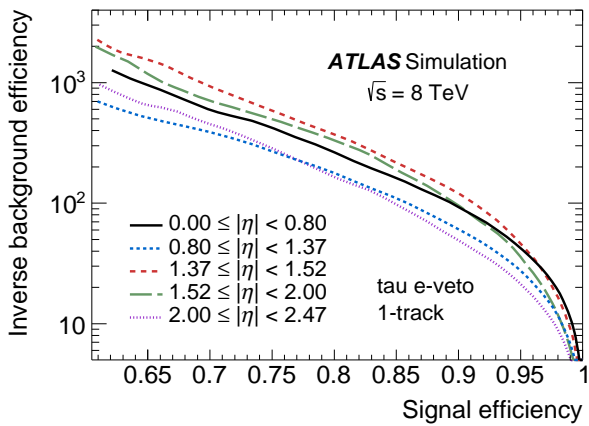}
\caption{
Electron veto inverse background efficiency versus signal efficiency in simulated samples, for 1-track \tauhadvis candidates. The background efficiency is 
determined using simulated $Z \to ee$ events.
}\label{fig:leptonveto_roc}
\end{center}
\end{figure}

\paragraph{Electron veto}
The characteristic signature of 1-track \tauhadvis can be mimicked by electrons. This creates a significant background contribution after all the
jet-related backgrounds are suppressed via kinematic, topological and \tauhadvis identification criteria. Despite the similarities of the \tauhadvis and electron
signatures, there are several properties that can be used to discriminate between them: transition radiation, which is more likely to be emitted by an
electron and causes a higher ratio $f_{\mathrm{HT}}$ of high-threshold to low-threshold track hits in the TRT for an electron than for a pion; 
the angular distance of the track from the \tauhadvis calorimeter-based direction; the ratio $f_{\mathrm{EM}}$ of energy deposited in the EM calorimeter
to energy deposited in the EM and HAD calorimeters; the amount of energy leaking into the hadronic calorimeter (longitudinal shower information) 
and the ratio of energy deposited in the region $0.1<\Delta R <0.2$ to the total core region $\Delta R<0.2$ (transverse shower information).
The distributions for two of the most powerful discriminating variables are shown in Fig.~\ref{fig:evetovars}.
These properties are used to define a \tauhadvis identification algorithm specialized in the rejection of electrons misidentified as hadronically decaying tau leptons, using a BDT. 
The performance of this \eveto algorithm is shown in Fig.~\ref{fig:leptonveto_roc}. Slightly different sets of variables are used in different $\eta$ regions. One of the 
reasons for this is that the variable associated with transition radiation (the leading track's ratio of high-threshold TRT hits to low-threshold TRT hits) is not available for $|\eta|>$ 2.0.
Three working points, labelled {\it tight, medium} and {\it loose} are chosen to yield signal efficiencies of 75\%, 85\%, and 95\%, respectively.

\paragraph{Muon veto}
Tau candidates corresponding to muons can in general be discarded based on the standard muon identification algorithms~\cite{muonPaper}.
The remaining contamination level can typically be reduced to a negligible level by a cut-based selection using the following characteristics.
Muons are unlikely to deposit enough energy in the calorimeters to be reconstructed as \tauhadvis candidates.
However, when a sufficiently energetic cluster in the calorimeter is associated with a muon, the muon track and the calorimeter cluster together may be misidentified as a \tauhadvis. Muons which deposit a large amount of energy in the calorimeter and therefore fail muon spectrometer reconstruction are characterized by a low electromagnetic energy fraction and a large ratio of track-\pT\ to \ET\ deposited in the calorimeter. 
Low-momentum muons which stop in the calorimeter and overlap with calorimeter deposits of different origin are characterized by a large electromagnetic energy fraction and a low \pT-to-\ET\ ratio.
A simple cut-based selection based on these two variables reduces the muon contamination to a negligible level.
The resulting efficiency is better than 96\% for true \tauhadvis, with a reduction of muons misidentified as \tauhadvis of about 40\%. However, the performance can vary depending on the \tauhadvis and muon identification levels.

\section{Efficiency measurements using $Z$ tag-and-probe data}\label{sec:performance}
To perform physics analyses it is important to measure the efficiency of the reconstruction and identification algorithms used online and offline 
with collision data.
For the \tauhadvis signal, 
this is done on a sample enriched in $Z \rightarrow \tau \tau$ events. For electrons misidentified as a tau signal 
(after applying the electron veto) this is done on a sample enriched in $Z \rightarrow ee$ events.

The chosen {\it tag-and-probe} approach consists of selecting events triggered by the presence of a lepton ({\it tag}) and containing a hadronically decaying tau lepton candidate ({\it probe}) in the final state and extracting the efficiencies 
directly from the number of reconstructed \tauhadvis before and after \tauid algorithms are applied. In practice, it is impossible to obtain a pure sample 
of hadronically decaying tau leptons, or electrons misidentified as a tau signal, and therefore backgrounds have to be taken into account. 
This is described in the following sections.

\subsection{Offline tau identification efficiency measurement}\label{sec:effoffline}
To estimate the number of background events for the purpose of tau identification efficiency measurements, a variable
with high separation power, which is modelled well for simulated \tauhadvis decays is chosen: the sum of the number of
core and outer tracks associated to the \tauhadvis candidate.  Outer tracks in $0.2 < \Delta R < 0.6$ are only considered
if they fulfill the requirement $D^\mathrm{outer} = min([~\pt^\mathrm{core} / \pt^\mathrm{outer}~] \cdot \Delta R(\mathrm{core,outer})) <$ 4,
where $\pt^\mathrm{core}$ refers to any track in the core region, and $\Delta R(\mathrm{core,outer})$ refers to the distance
between the candidate outer track and any track in the core region.  This requirement suppresses the contribution of outer
tracks from underlying and pile-up events, due to requirements on the relative momentum and separation of the tracks.  This
allows the signal track multiplicity to retain the same structure as the core track multiplicity distribution.  For backgrounds
from multi-jet events, the track multiplicity is increased by the addition of tracks with significant momentum in the outer cone.
The requirement on $D^\mathrm{outer}$ was chosen to offer optimal signal to background separation.
A fit is then performed using the expected distributions of this variable for both signal and background to 
extract the \tauhadvis signal. This fit is performed for each exclusive \tauid working point, 
corresponding to: candidates failing the {\it loose} requirement, candidates satisfying the {\it loose} requirement but failing the {\it medium} requirement,
candidates satisfying the {\it medium} requirement but failing the {\it tight} requirement and candidates satisfying the {\it tight} requirement.

\subsubsection{Event selection}
$Z\rightarrow\taulep\tauhad$ events are selected by a triggered muon or electron coming from the leptonic decay of a tau lepton, 
and the hadronically decaying tau lepton is then searched for in the rest of the event, considered as the {\it probe} for the tau identification performance measurement.
These events are triggered by a single-muon or a single-electron trigger 
requiring one isolated trigger muon or electron with a $p_{\rm T}$ of at least 24 \GeV.

Offline, muons and electrons with $p_{\rm T}>26$ \GeV\ are thereafter selected, representing the {\it tag} objects. 
Additional track and calorimeter isolation requirements are applied to the muon and electron.
Identified muons are required to have $|\eta|<2.4$. Identified electrons are required to have $|\eta|<1.37$ or $1.52<|\eta|<2.47$, therefore excluding the poorly instrumented region at the interface between the barrel and endcap calorimeters.  In addition to the requirement of exactly one isolated muon or electron ($\ell$), a \tauhadvis candidate is selected in the kinematic range $p_{\rm T} > 15$\,\GeV{} and $|\eta|<2.5$, requiring one or three associated tracks in the core region 
and an absolute electric charge of one and no geometrical overlap with muons with $\pt{} > 4$\,\GeV{} or with electrons with $\pt{} > 15$\,\GeV{} of {\it loose} or {\it medium} quality (depending on $\eta$).  For \tauhadvis with one associated track, a muon veto and a {\it medium} \eveto is applied. 
In addition to this, a very loose requirement on the \tauid BDT score is made 
which strongly suppresses jets while being more than 99\% efficient for $Z \to \tau\tau$ signal. 
The tag and the probe objects are required to have opposite-sign electric charges (OS).

Additional requirements are made in order to suppress $(Z \rightarrow \ell \ell)$ + jets and $(W\rightarrow \ell \nu_{\ell})$ + jets events:
\begin{itemize}

\item On the invariant mass calculated from the lepton and the \tauhadvis four-momenta ($m_{\rm vis}(\ell,\tauhadvis)$): 
for $\pt^{\tauhadvis}<20 \GeV$, $45 \GeV <m_{\rm vis}( \ell,\tauhadvis)< 80 \GeV$. Otherwise, for the $\mu$ channel, $50 \GeV <m_{\rm vis}(\mu,\tauhadvis)< 85 \GeV$, 
and for the $e$ channel: $50 \GeV <m_{\rm vis}( e ,\tauhadvis)< 80 \GeV$.
For the signal, this variable peaks in these regions.

\item On the transverse mass of the lepton and \met{} system ($m_{\rm T}=\sqrt{2p^{\ell}_{\rm T} \cdot E^{\rm miss}_{\rm T}(1-\cos\Delta\phi(\ell,E^{\rm miss}_{\rm T}))}$):
$m_{\rm T} <$ 50 GeV. 
For most backgrounds (e.g. $(W\rightarrow~\ell\nu_{\ell})$~+~jets), this variable peaks at larger values. 

\item On the distance in the azimuthal plane between the lepton and \met{} (neutrinos) and between the \tauhadvis and \met{} ($\sumcosdphi=\cos\Delta\phi(\ell, E^{\rm miss}_{\rm T})+\cos\Delta\phi(\tauhadvis , E^{\rm miss}_{\rm T})$): $\sumcosdphi > -0.15$.  For the signal, this variable tends to peak at zero, indicating that the neutrinos point mainly in the direction of one of the two leptons from $Z$ decay products.  For $W$ + jets background events, the value is typically negative, indicating that the neutrino points away from the two lepton candidates.
\end{itemize}

\subsubsection{Background estimates and templates}
\label{sec:backback}
The signal track multiplicity distribution is modelled using simulated $Z\rightarrow\tau_{\mathrm{lep}}\tauhad$ events. Only reconstructed \tauhadvis 
matched to a true hadronic tau decay are considered. 

A single template is used to model the background from quark- and gluon-initiated jets that are misidentified as hadronic tau decays. The background is mainly composed of multi-jet and $W$+jets events with a minor contribution from $Z$+jets events. The template is constructed starting from a enriched multi-jet control region in data that uses the full signal region selection but requires that the tag and probe objects have same-sign charges (SS).
The contributions from $W$+jets and $Z$+jets in the SS control region are subtracted. The template is then scaled by the ratio of OS$/$SS multi-jet events, measured in a control region which inverts the very loose identification requirement of the signal region. Finally, the OS contributions from $W$+jets and $Z$+jets are added to complete the template. The $Z$+jets contribution is estimated using simulated samples. The shape of the $W$+jets contribution is estimated from a high-purity $W$+jets control region, defined by removing the $m_{\mathrm{T}}$ requirement and inverting the requirement on $\sumcosdphi$. The normalization of the $W$+jets contribution is estimated using simulation.

An additional background shape is used to take into account the contamination due to misidentified electrons or muons.
This small background contribution (stemming mainly from $Z\rightarrow \ell \ell $ events) is modelled by taking the shape predicted by simulation 
using candidates which are not matched to true \tauhadvis\ in events of type $Z\rightarrow\tau_{\mathrm{lep}}\tauhad$, $t\bar{t}$, diboson, $Z\rightarrow ee,\mu\mu$ where 
the reconstructed tau candidate probe is matched to a electron or muon. For the fit, the contribution of these backgrounds is fixed to the value predicted by the simulation,
which is typically less than 5\% of the total signal yield.

To measure both the 1-track and 3-tracks \tauhadvis\ efficiencies, a fit of the data to the model (signal plus background) is performed, using two separate signal templates. 
The signal templates are obtained by requiring exactly one or three tracks reconstructed in the core region of the \tauhadvis candidate.
To improve the fit stability in the background-dominated region where the tau candidates fail the {\it loose} requirements, the ratio of the 1-track to 3-track normalization is fixed to the value predicted by the simulation. For other exclusive regions, the ratio is allowed to vary during the fit. 

In the fit to extract the efficiencies for real tau leptons passing different levels of identification, the ratio of jet to other \tauhadvis candidates is  determined in a preselection step (where no identification is required) and then extrapolated to regions where identification is required by using jet misidentification rates determined in an independent data sample.

\subsubsection{Results}
\begin{figure}[htbp]
\begin{center}
\subfigure[]{
        \includegraphics[width=0.47\textwidth]{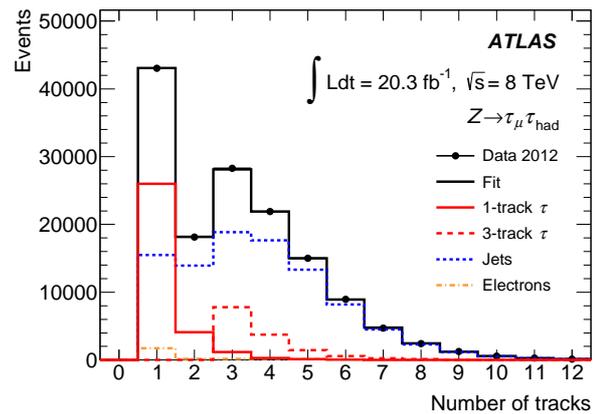}
}
\subfigure[]{
        \includegraphics[width=0.47\textwidth]{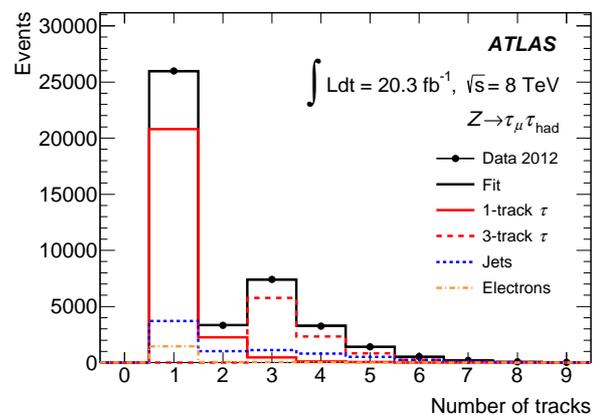}
}
\caption{
Template fit result in the muon channel, inclusive in $\eta$ and \pt\ for $\pt>20$ \,\GeV\ for the offline \tauhadvis candidates (a) before the requirement of \tauid, and (b) fulfilling the {\it medium} \tauid requirement.
}\label{fig:template_fit2}
\end{center}
\end{figure}

Figure~\ref{fig:template_fit2} shows an example of the track multiplicity distribution after the tag-and-probe selection, before and after applying the \tauid
requirements, with the results of the fit performed.  The peaks in the one- and three-track bins are due to the signal contribution.  These are visible
before any identification requirements are applied, and become considerably more prominent after identification requirements are applied, due to the large
amount of background rejection provided by the identification algorithm.
To account for the small differences between data and simulations, correction factors, defined as the ratio of the efficiency in data to the efficiency in simulation for \tauhadvis\ signal to pass a certain level of identification, are derived. 
Their values are compatible with one, except for the {\it tight} 1-track working point, where the correction factor is about 0.9.

\begin{figure}[htbp]
\begin{center}

\subfigure[]{
  \includegraphics[width=0.47\textwidth]{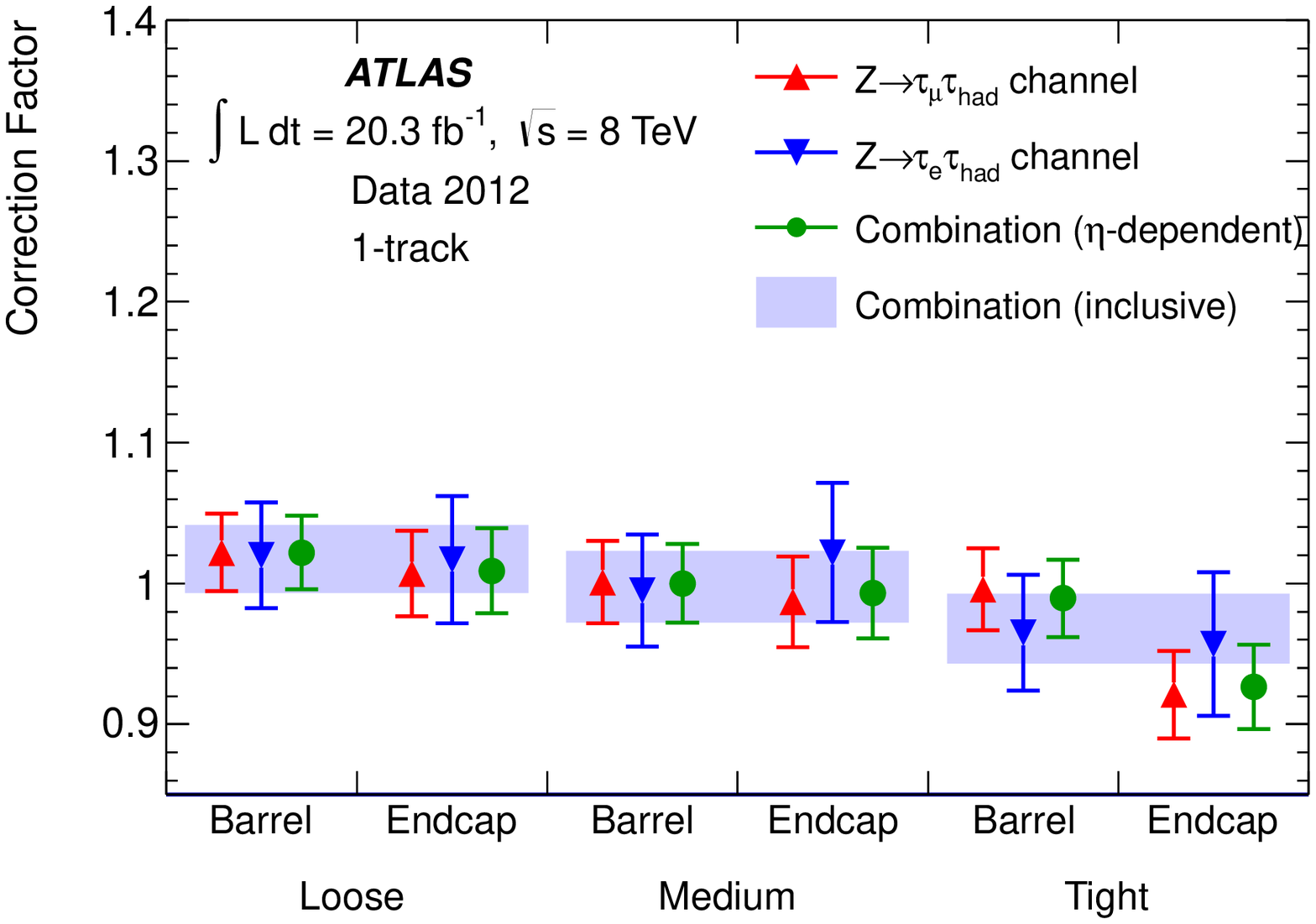}
}
\subfigure[]{
  \includegraphics[width=0.47\textwidth]{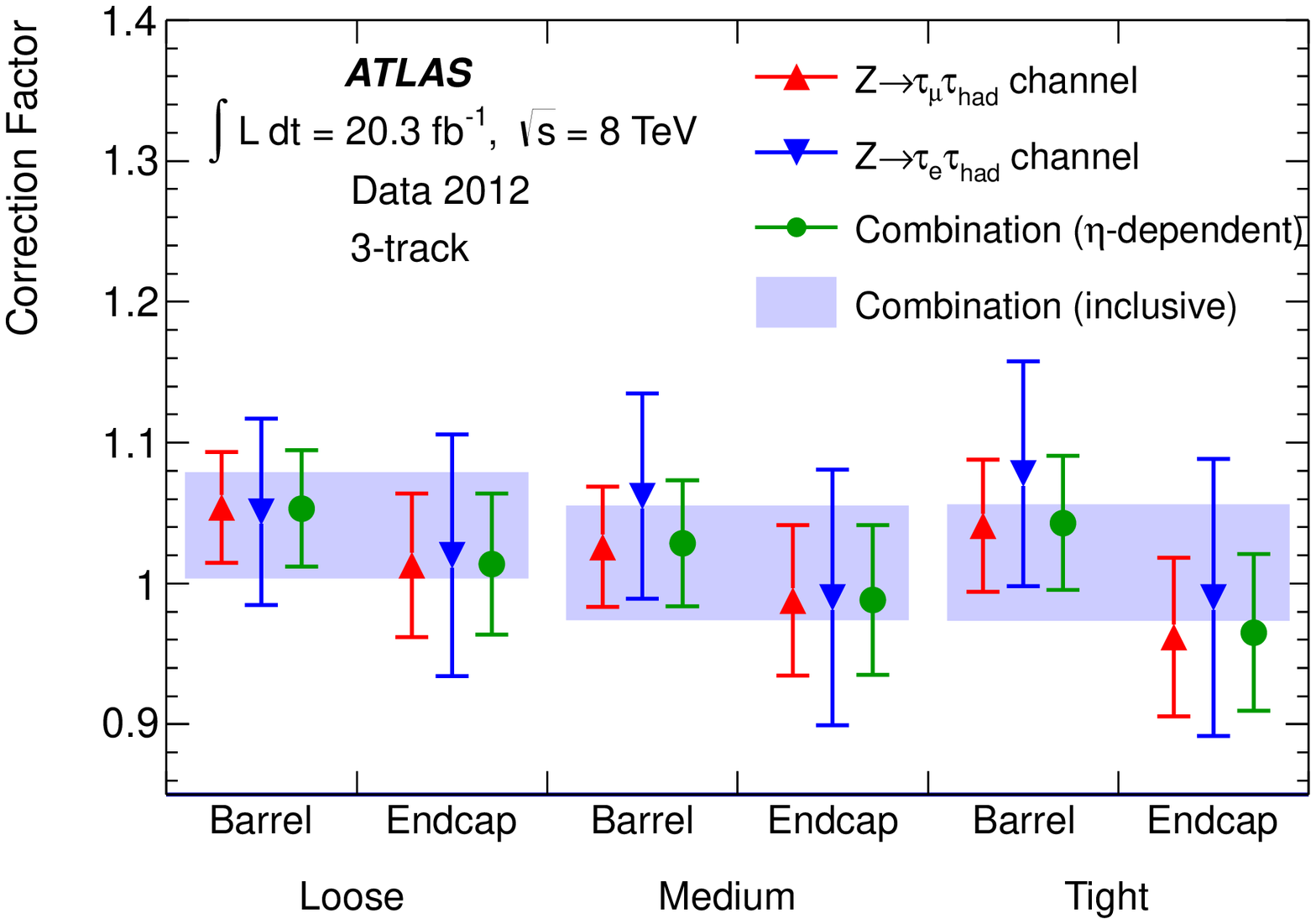}
}

\caption{
Correction factors needed to bring the offline tau identification efficiency in simulation to the level 
observed in data, for all \tauid working points as a function of $\eta$. The combinations of the muon and electron channels are also shown, and the results are displayed separately for (a) 1-track and (b) 3-track \tauhadvis candidates with $\pt{} > 20$ \GeV.
The combined systematic and statistical uncertainties are shown.
}\label{fig:template_sf}
\end{center}
\end{figure}

Results from the electron- and muon-tag analysis are combined to improve the precision of the correction factors, shown in Fig.~\ref{fig:template_sf}.
No significant dependency on the \pt{} of the \tauhadvis is observed and hence the results are provided separately only for the barrel ($|\eta| < 1.5$) and the
endcap ($1.5 < |\eta| < 2.5$) region, and for one and three associated tracks.
Uncertainties depend slightly on the \tauid level and kinematic quantities. In Table~\ref{tab:tauid_error}, the most important systematic uncertainties for the working point used by
most analyses,  {\it medium} \tauid, are shown, together with the total statistical and systematic uncertainty. 
Uncertainties due to the underlying event (UE) are the dominant ones for the signal template, and are estimated by comparing {\sc alpgen-Herwig} and {\sc Pythia} simulations. The shower
model and the amount of detector material are also varied and included in the number reported in Table~\ref{tab:tauid_error}. The $W$+jets shape uncertainty accounts for differences
between the $W$+jets shape in the signal and control regions and is derived from comparisons to simulated $W$+jets events. The jet background fraction uncertainty accounts
for the effect of propagating the statistical uncertainty on the jet misidentification rates.

The results apply to \tauhadvis candidates with $\pt{} > 20$ \GeV. For $\pt{} < $ 20 \GeV, uncertainties increase to a maximum of 15\% for inclusive \tauhadvis candidates.  For $\pt{} >$ 100 GeV, there are no abundant sources of hadronic tau decays to allow for an efficiency measurement. Previous studies using high-$\pt{}$ dijet events indicate that there is no degradation in the modelling of tau identification in this $\pt{}$ range, within the statistical uncertainty of the measurement~\cite{Aad:2012gm}.

\begin{table}[htp]

\begin{center}
\begin{tabular}{lcc}
\hline\hline\noalign{\smallskip}
Source       & \multicolumn{2}{c}{Uncertainty [\%]} \\
                  & 1-track & 3-track \\
\noalign{\smallskip}\hline\noalign{\smallskip}
Jet background fraction   &     0.8 &      1.5 \\
Jet template shape  &     0.9 &      1.4 \\
Tau energy scale     &     0.7 &      0.8 \\
Shower model/UE   &     1.8 &      2.5 \\
Statistics        &     1.0 &      2.2 \\
\noalign{\smallskip}\hline\noalign{\smallskip}
Total             &     2.5 &      4.0 \\
\hline\hline\noalign{\smallskip}
\end{tabular}
\end{center}
\caption{
Dominant uncertainties on the {\it medium} \tauid efficiency correction factors estimated with the $Z$ tag-and-probe method, and the total uncertainty, which combines systematic and statistical uncertainties.  These uncertainties apply to \tauhadvis candidates with \pt{} $>$ 20 GeV.
\label{tab:tauid_error}
}
\end{table}

\subsection{Trigger efficiency measurement}\label{sec:triggerperf}
The tau trigger efficiency is measured with \Ztau\ events using tag-and-probe selection similar to the one
described in Sect.~\ref{sec:effoffline}. The only difference is that the efficiency is measured 
with respect to identified offline \tauhadvis candidates and thus, offline \tauid selection criteria are applied during the event
selection. Only the muon channel is considered, as the background contamination is smaller than in the 
electron channel. The statistical uncertainty improvements that could be obtained by the addition of
the electron channel are offset by the larger systematic uncertainties associated with this
channel.
The systematic uncertainties are also different from those in the offline identification measurement, since 
the purity after identification is already high.  The systematics are dominated by the uncertainties on
the modelling of the kinematics of the background events, rather than the total normalization, as is the case for the
offline identification measurement.

The dominant background contribution is due to \Wboson\ + jets and multi-jet events, where a jet is misidentified as a \tauhadvis. These backgrounds are 
estimated using a method similar to the one described in Sect.~\ref{sec:backback}. 
The same multi-jet and \Wboson\ + jets control regions are used.
The shape of other backgrounds is taken from simulation but the normalizations of the dominant backgrounds are estimated from data control regions.
The contribution of top quark events is normalized in a control region requiring one jet originating from a $b$-quark. $Z$+jets events with leptonic $Z$ decays and one
of the additional jets being misidentified as \tauhadvis are normalized by measuring this misidentification rate in a control region with two identified oppositely charged same-flavour
leptons.

In total, more than 60,000 events are collected, with a purity of about 80\% when the offline {\it medium} \tauid
requirement is applied. With the addition of the tau trigger requirement, the purity increases to about 88\%.
Most of the backgrounds accumulate in the region $\pt<30$\,\GeV. 

\begin{figure}[htbp]
\begin{center}
\subfigure[]{
\includegraphics[width=0.47\textwidth]{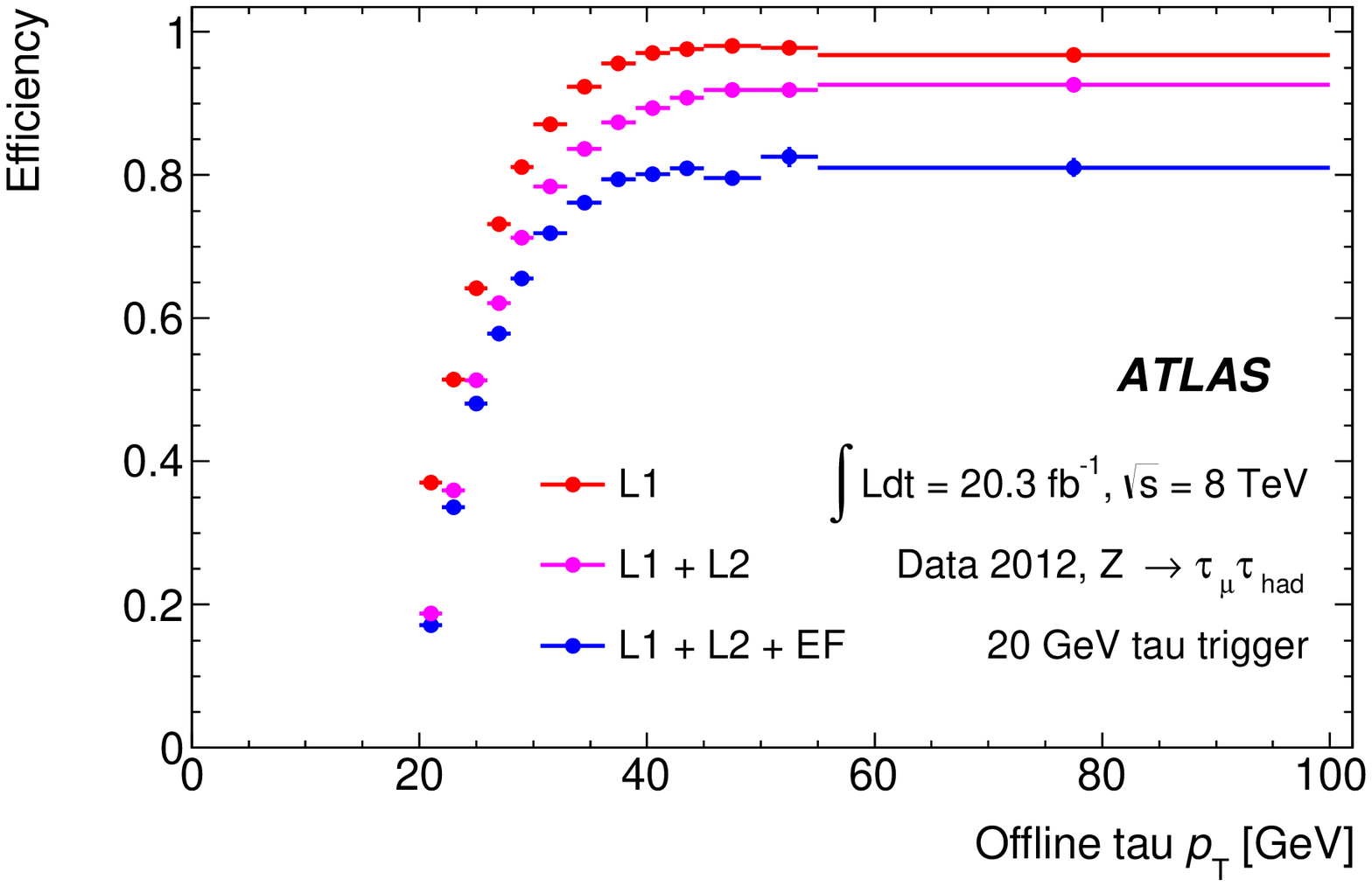}
}
\subfigure[]{
\includegraphics[width=0.47\textwidth]{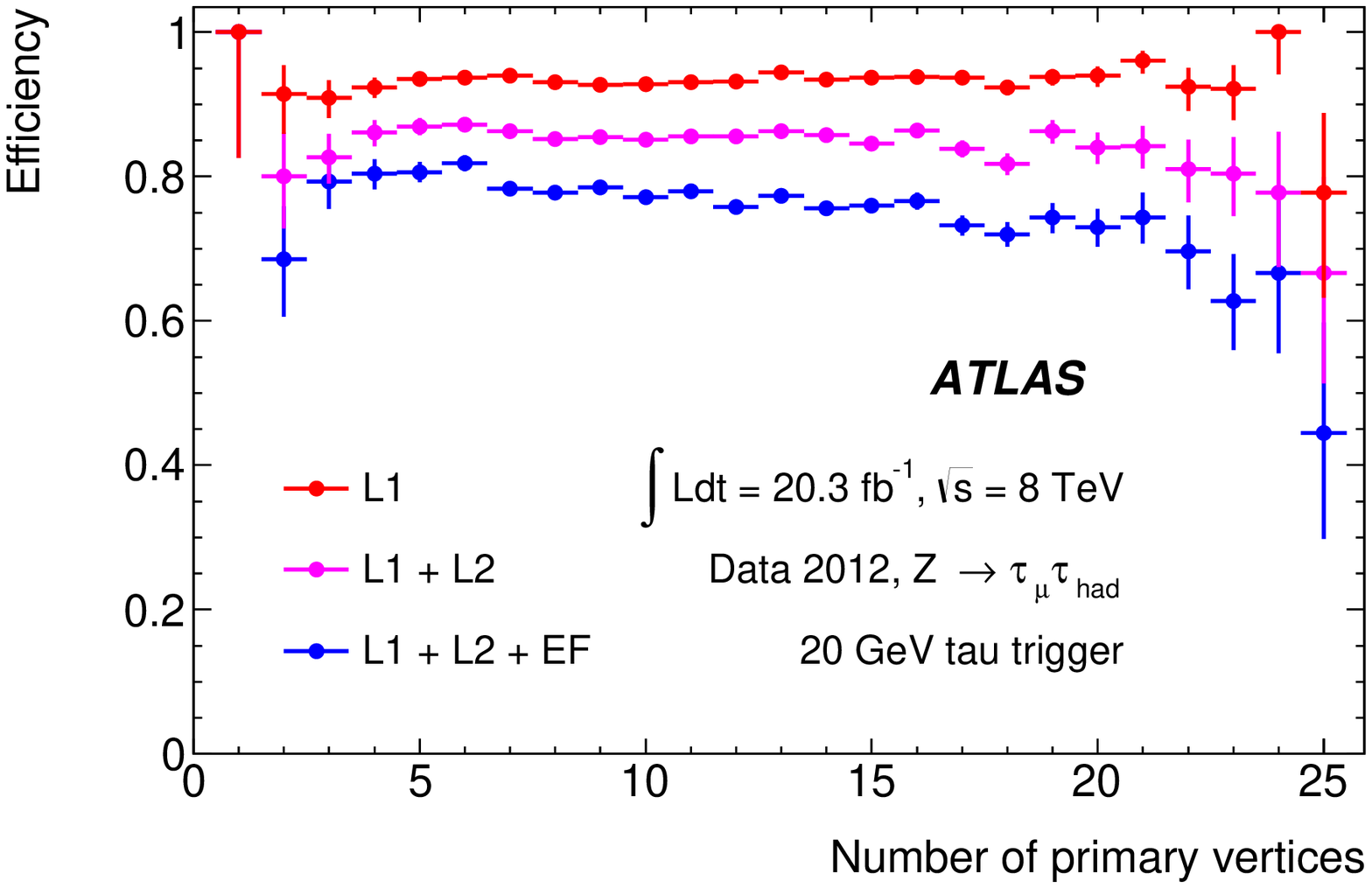}
}
\caption{
The tau trigger efficiency for \tauhadvis candidates identified by the offline $medium$ tau identification, as a function 
of (a) the offline \tauhadvis transverse energy and (b) the number of primary vertices.
The error bars correspond to the statistical uncertainty in the efficiency.
}
\label{fig:trigL1L2EF}
\end{center}
\end{figure}

\begin{figure}[htbp]
\begin{center}
\includegraphics[width=0.49\textwidth, height=6.2cm]{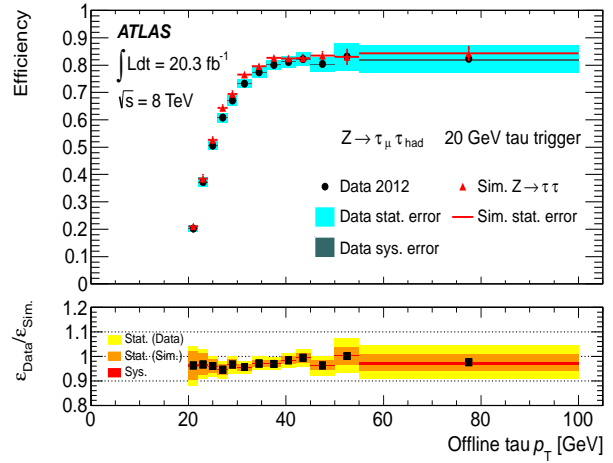}
\caption{
The measured tau trigger efficiency in data and simulation, for the offline \tauhadvis candidates passing 
the {\it medium} \tauid, as a function of offline \tauhadvis transverse energy. 
The expected background contribution has been subtracted from the data. 
The uncertainty band on the ratio reflects the statistical uncertainties associated with
data and simulation and the systematic uncertainty associated with the background subtraction in data. 
}
\label{fig:trigeff}
\end{center}
\end{figure}

Figure~\ref{fig:trigL1L2EF} shows the measured tau trigger efficiency for \tauhadvis candidates identified by the offline $medium$ tau identification
as functions of the offline \tauhadvis transverse energy and the number of primary vertices in 
the event, for each level of the trigger. 
The tau trigger considered has calorimetric isolation and a \pt\ threshold of 11 \GeV\ at L1, 
a 20 \GeV\ requirement on \pt{}, the number of tracks restricted to three or less, and $medium$ selection on the BDT score at EF. 
The efficiency depends minimally on $\pt$ for $\pt>35$\,\GeV\ or on the pile-up conditions. 
The measured tau trigger efficiency is compared to simulation in Fig.~\ref{fig:trigeff}; the efficiency
is shown to be modelled well in simulation. Correction factors, as defined in Sect.~\ref{sec:effoffline},
are derived from this measurement. The correction factors are in general compatible with unity, except for the region
$\pt<40$\,\GeV\ where a difference of a few per cent is observed.

In the \pt{} range from 30 \GeV\ to 50 \GeV, the uncertainty on the correction factors is about 2\% but increases to about 8\% for $\pt=100$\,\GeV. 
The uncertainty is also sizeable in the region $\pt<30$\,\GeV, where the background contamination is the largest.

\subsection{Electron veto efficiency measurement}
\begin{figure}[htbp]
\begin{center}
\subfigure[]{
        \includegraphics[width=0.47\textwidth]{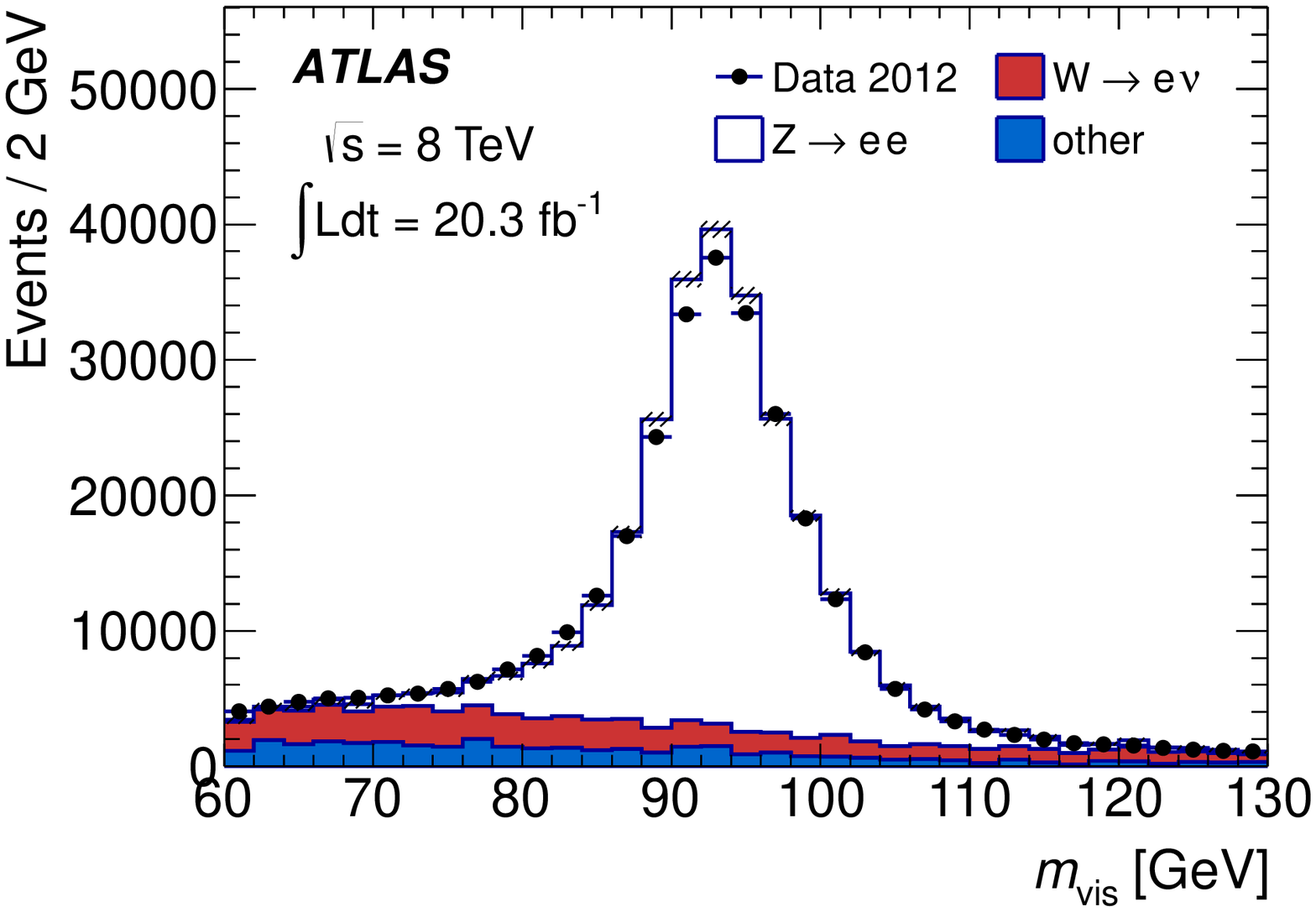}
}
\subfigure[]{
        \includegraphics[width=0.47\textwidth]{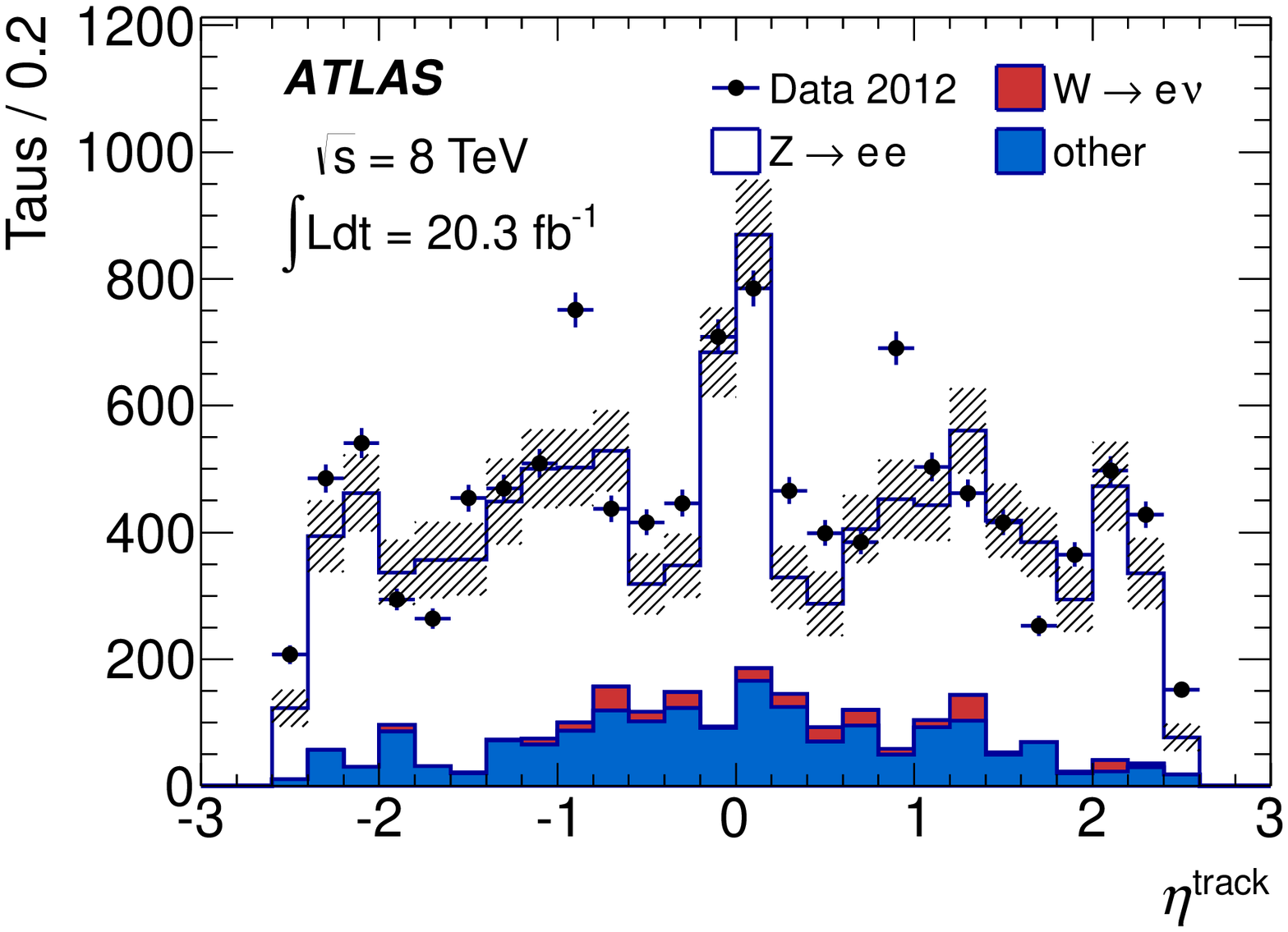}
}
\caption{
(a) Visible mass of electron--positron pairs for the offline \eveto efficiency measurement, after tag-and-probe selection, where the probe lepton passes {\it medium} \tauid and does 
not overlap with {\it loose} electrons, before the \eveto is applied. (b) $\eta$ distribution for \tauhadvis candidates (electrons misidentified as hadronic tau decays) after applying a {\it loose} \eveto. Uncertainties shown are only statistical.
}\label{fig:leptonveto_sf}
\end{center}
\end{figure}

To measure the efficiency for electrons reconstructed as \tauhadvis to pass the electron veto in data, a tag-and-probe analysis singles out a pure sample of $Z \rightarrow ee$ events, 
as illustrated in Fig.~\ref{fig:leptonveto_sf} (a). The measurement uses
probe 1-track \tauhadvis candidates in the opposite hemisphere to the identified tag electron.  The tag electron is required to fulfil $p_{\mathrm{T}}^{\mathrm{tag}} > 35$ GeV in order to suppress backgrounds from $Z \rightarrow \tau\tau$ events. The probe is required not to overlap geometrically 
with an identified electron, e.g. in the case of Fig.~\ref{fig:leptonveto_sf} a {\it loose} electron identification is used. Different veto algorithms are tested in combination with different levels of jet discrimination, and the effects estimated.
Efficiencies are extracted directly from the number of reconstructed \tauhadvis before and after identification, in bins of $\eta$ of the \tauhadvis 
candidate, after subtracting the background modelled by simulation (normalized to data in dedicated control regions).
The shape and normalization of the multi-jet background distribution for the $\eta$ of the \tauhadvis are estimated using events with SS tag electron and probe \tauhadvis in data after subtracting backgrounds in the SS region using simulation.  To estimate the $W\rightarrow e\nu$, $Z\rightarrow\tau\tau$, and $t\bar{t}$ backgrounds, the shape of this distribution is obtained from simulation but normalized to dedicated data control regions 
for each background.

Differences in the modelling of the \eveto algorithm's performance in simulation compared to data are parameterized as correction factors in bins of $\eta$ of the \tauhadvis 
candidate, by comparing distributions similar to the one shown in Fig.~\ref{fig:leptonveto_sf} (b). 

Uncertainties on the correction factors (which are typically close to unity) are $\eta$-dependent and amount to about 10\% for the {\it loose} \eveto and get larger for the {\it medium} and {\it tight} \eveto working points, mainly driven by statistical uncertainties. 
A summary of the main uncertainties for the working point shown in Fig.~\ref{fig:leptonveto_sf} is provided in Table~\ref{tab:eveto_error}.

\begin{table}[htp]
\begin{center}
\begin{tabular}{lc}
\hline\hline\noalign{\smallskip}
Source & Uncertainty [\%] \\
\noalign{\smallskip}\hline\noalign{\smallskip}
Tag selection (\pt{}, isolation) & 5--28 \\
Background rejection             & 1--8  \\
Statistics                       & 7--12 \\
\noalign{\smallskip}\hline\noalign{\smallskip}
Total                            & 8--30 \\
\hline\hline\noalign{\smallskip}
\end{tabular}
\end{center}
\caption{
Dominant uncertainties on the {\it loose} \eveto efficiency correction factors estimated with the $Z$ tag-and-probe method. 
The range of the uncertainties reflects their variation with $\eta$.
\label{tab:eveto_error}
}
\end{table}

\section{Calibration of the \tauhadvis energy}\label{sec:tes}
The \tauhadvis energy calibration is done in several steps.
First, a calibration described in Sect.~\ref{sec:tesmc} and derived from simulation brings 
the tau energy scale (TES) into agreement with the true energy scale 
at the level of a few per cent and removes 
any significant dependencies of the energy scale on the pseudorapidity, energy, pile-up conditions 
and track multiplicity. Then, additional small corrections to the TES are 
derived using one of two independent data-driven methods described in Sect.~\ref{sec:offlineres}. 
Which of the two methods is used depends on whether for a given study the agreement 
between reconstructed and true TES or the modelling of the TES in simulation is more important.

\subsection{Offline \tauhadvis energy calibration}\label{sec:tesmc}
The clusters associated with the \tauhadvis reconstruction are calibrated at the LC scale. 
For anti-$k_t$ jets with a distance parameter $R=0.4$, this calibration accounts for the 
non-compensating nature of the ATLAS calorimeters and for energy deposited outside the reconstructed 
clusters and in non-sensitive regions of the calorimeters. However, it is neither optimized 
for the cone size used to measure the \tauhadvis momentum ($\Delta R=0.2$) nor
for the specific mix of hadrons observed in tau decays; and it does not correct for the underlying event or for pile-up contributions. 
Thus an additional correction is needed to obtain an energy scale 
which is in agreement with the true visible energy scale, thereby also improving the \tauhadvis energy resolution.

This correction (also referred to as a response curve) is computed as a function of $E^\tau_\mathrm{LC}$ using $Z\to\tau\tau$, $W \to \tau\nu$ and 
$Z' \to \tau\tau$ events simulated with {\sc Pythia8}. Only \tauhadvis candidates with reconstructed 
$\ET > 15$ \GeV\ and $\abseta<2.4$ matched to a true \tauhadvis with $E^\mathrm{true}_{\mathrm{T,vis}} >10$ \GeV\ are considered. 
Additionally, they are required to satisfy {\it medium} \tauid criteria and to have a distance $\Delta R>0.5$ to other reconstructed jets. 
The response is defined as the ratio of the reconstructed \tauhadvis energy at the LC 
scale $E^\tau_\mathrm{LC}$ to the true visible energy $E^\mathrm{true}_\mathrm{vis}$. 

The calibration is performed in two steps: 
first, the response curve is computed; then, additional small corrections for the 
pseudorapidity bias and for pile-up effects are derived.

\begin{figure}[htbp]
\begin{center}
\subfigure[]{
        \includegraphics[width=0.47\textwidth]{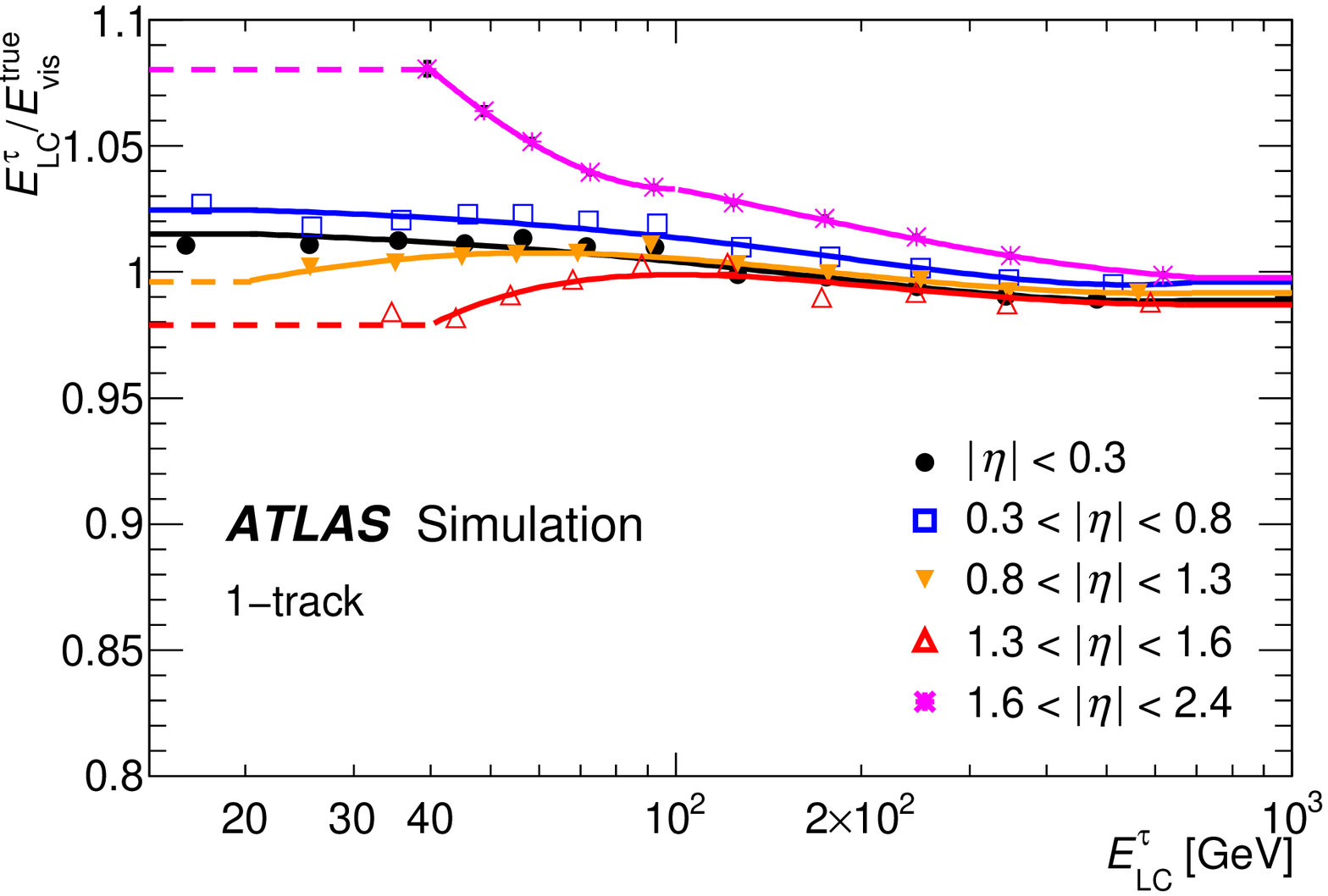}
}
\subfigure[]{
        \includegraphics[width=0.47\textwidth]{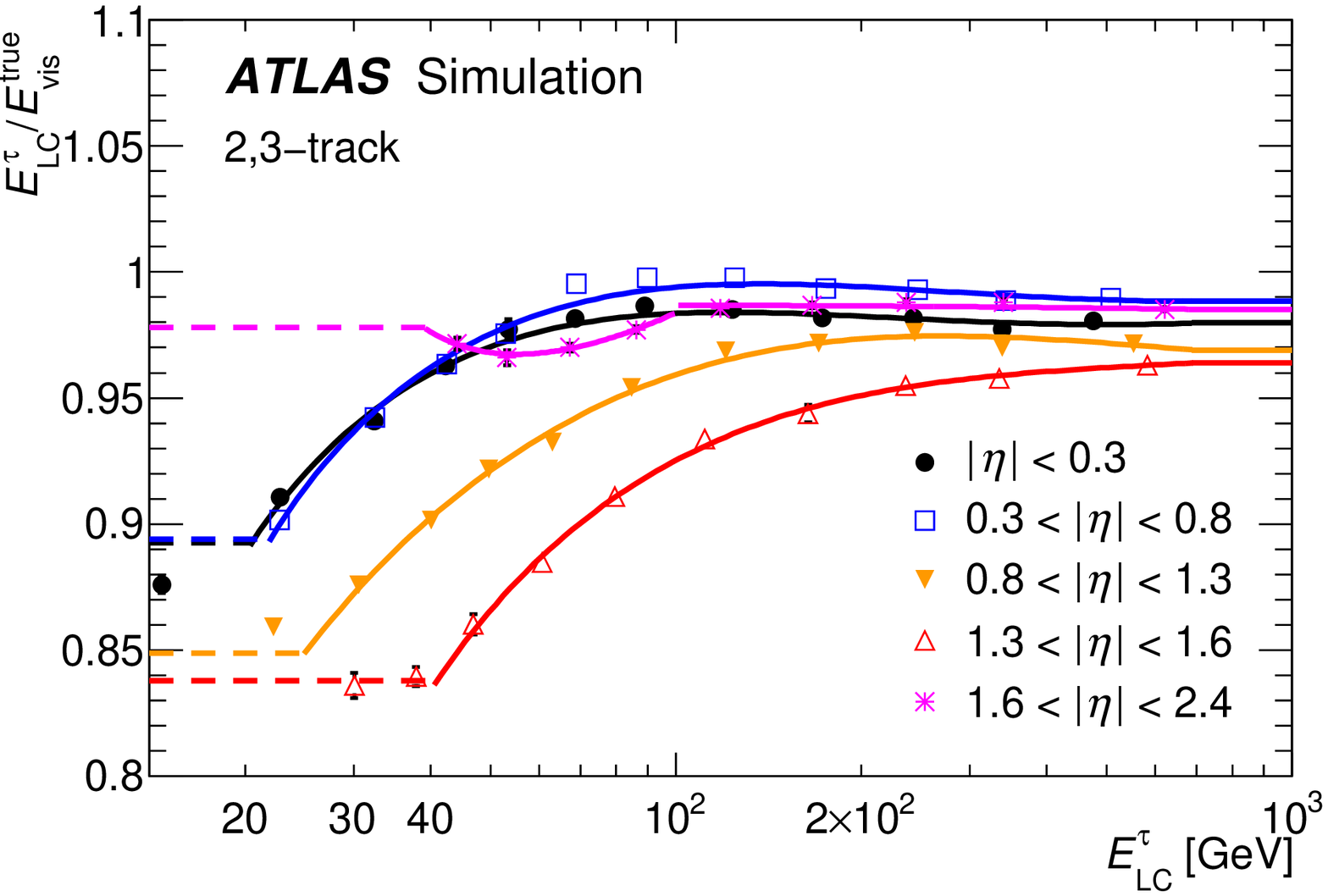}
}
\caption{
Offline \tauhadvis energy response curves as a function of the reconstructed \tauhadvis energy $E^\tau_\mathrm{LC}$ for hadronic tau decays with (a) one and (b) 
more than one associated tracks. One curve per pseudorapidity region 
$|\eta^\mathrm{LC}|$ is shown. The region where markers are shown corresponds 
approximately to a transverse energy $E^\tau_\mathrm{T,LC}>15$ \GeV. 
For very low and very high energies, 
the response curves are assumed to be constant.
Uncertainties are statistical only.
}\label{fig:tescalib}
\end{center}
\end{figure}

The response curve is evaluated in intervals of $E^\mathrm{true}_{\mathrm{vis}}$ and of the absolute value of 
the reconstructed \tauhadvis pseudorapidity for \tauhadvis candidates with one or more tracks. 
In each interval, the distribution of this ratio is fitted with a Gaussian function to determine the mean value. 
This mean value as a function of the average $E^\tau_\mathrm{LC}$ in a given interval is then 
fitted with an empirically derived functional form. The resulting functions are shown in Fig.~\ref{fig:tescalib}. 

After using this response curve to calibrate hadronically decaying tau leptons their reconstructed mean energy 
is within 2\% of the final scale, which is set using two additional small corrections. 
First, a pseudorapidity correction is applied, which is necessary to counter a bias due to 
underestimated reconstructed cluster energies in poorly instrumented regions. The correction 
depends only on $|\eta^\mathrm{LC}|$ and is smaller than 0.01 units in the 
transition region between the barrel and endcap electromagnetic calorimeters and negligible 
elsewhere, leading to the final reconstructed pseudorapidity $\eta^\mathrm{rec}=\eta^\mathrm{LC} - \eta^\mathrm{bias}$.

Pile-up causes response variations of typically a few per cent. This is corrected by subtracting 
an amount of energy which is proportional to the number of reconstructed proton--proton interaction vertices $n_\mathrm{vtx}$ 
in a given event. The parameter describing the proportionality is derived for different regions of 
$|\eta^\mathrm{rec}|$ using a linear fit versus $n_\mathrm{vtx}$, 
for \tauhadvis candidates with one or more tracks. The correction varies in the range 90--420 MeV per reconstructed vertex, increasing with $|\eta|$.

\begin{figure}[htbp]
\begin{center}
\subfigure[]{
        \includegraphics[width=0.47\textwidth]{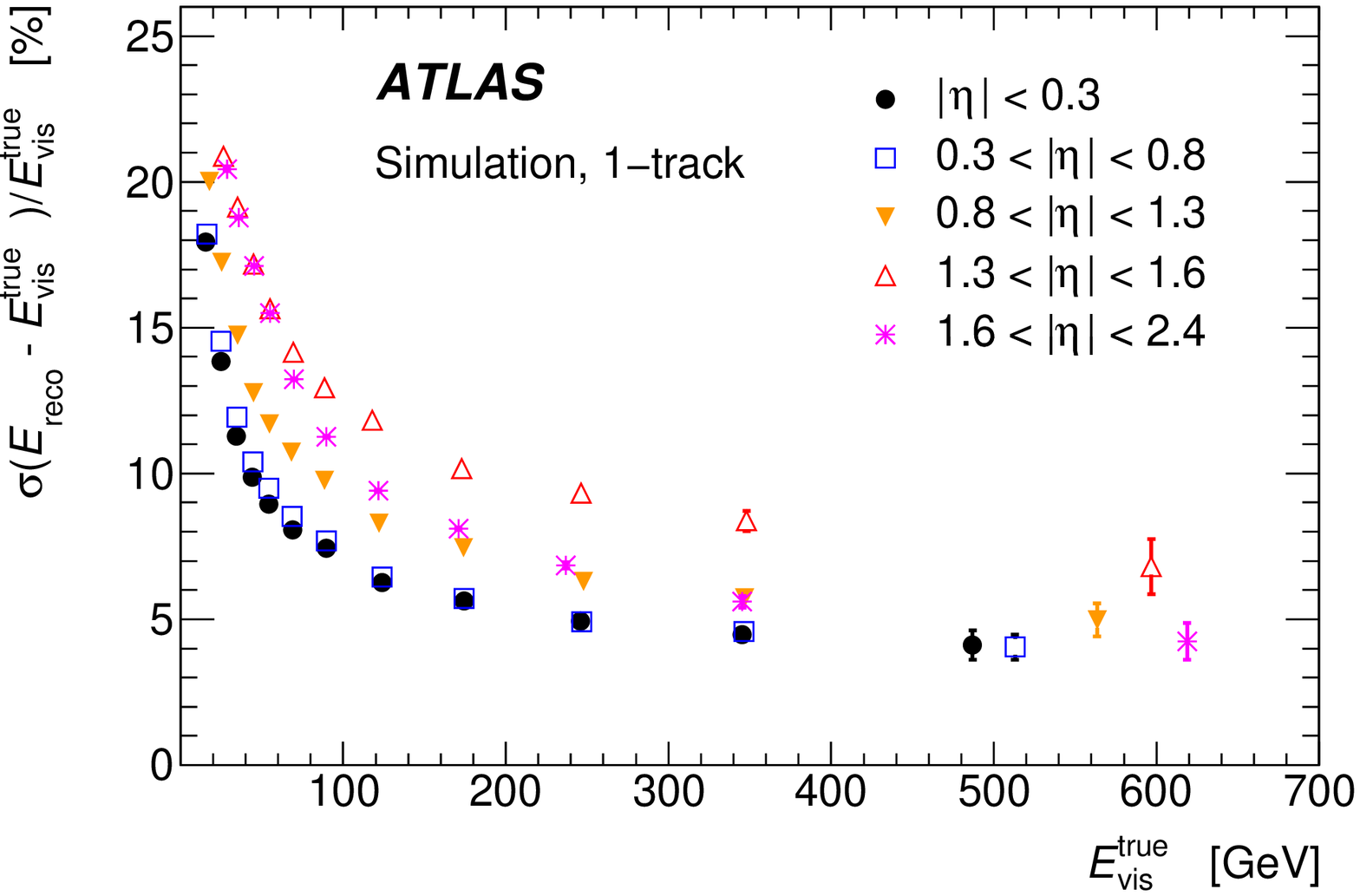}
}
\subfigure[]{
        \includegraphics[width=0.47\textwidth]{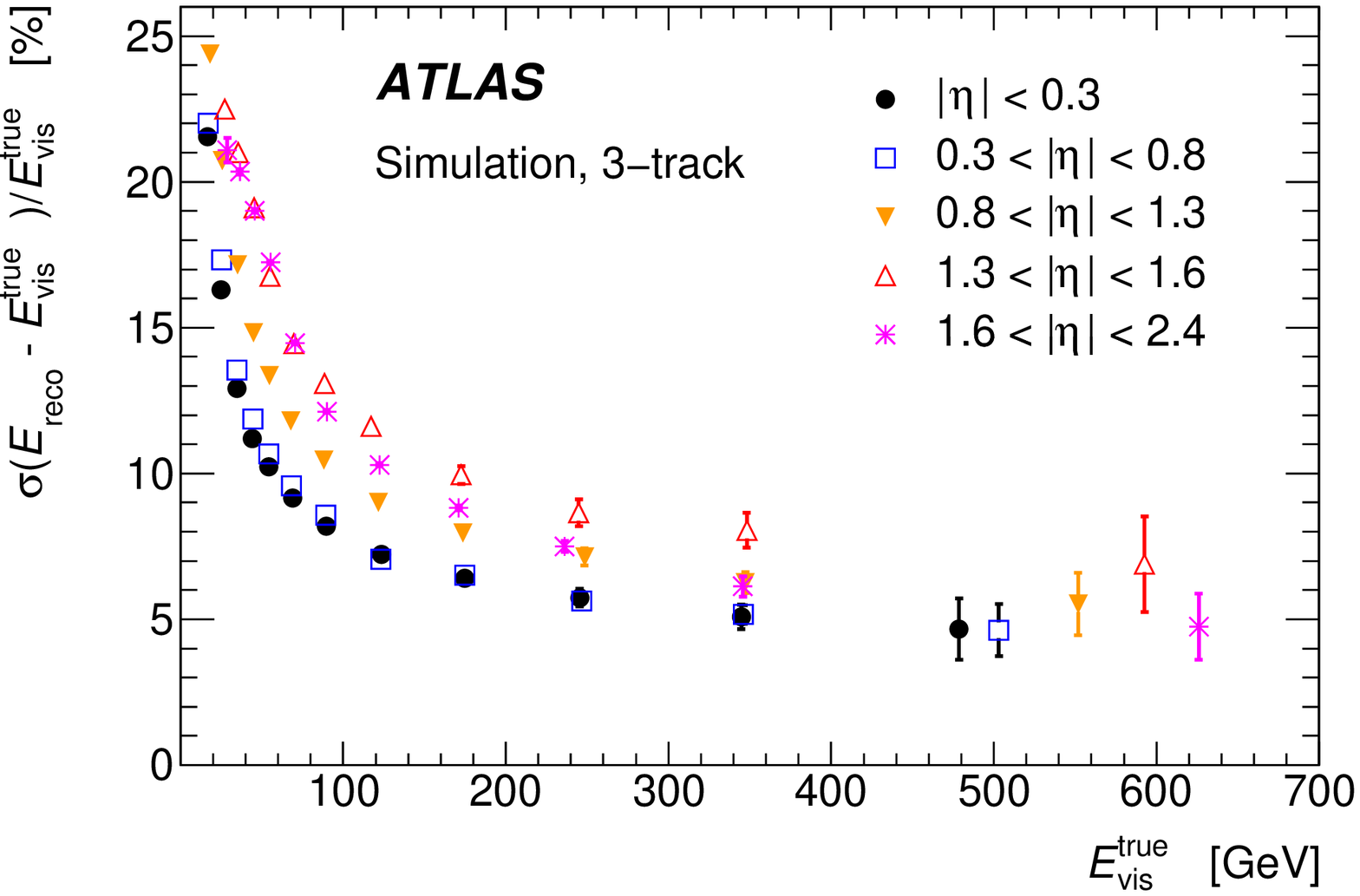}
}
\caption{
Offline energy resolution for hadronically decaying tau leptons, separately for (a) one and (b) three 
associated tracks and for different pseudorapidity regions. The resolution shown is the standard 
deviation of a Gaussian function fit to the distribution of
$(E_\mathrm{reco}-E^\mathrm{true}_\mathrm{vis})/E^\mathrm{true}_\mathrm{vis}$ in a given 
range of $E^\mathrm{true}_\mathrm{vis}$ and $|\eta^\mathrm{true}_\mathrm{vis}|$.
}\label{fig:resolution}
\end{center}
\end{figure}

The energy resolution, as determined from simulated data, as a function of the true visible
energy after the complete tau calibration is shown in Fig.~\ref{fig:resolution}. The resolution
is about 20\% at very low $E$ and reduces to about 5\% for energies above a few hundred \GeV. 
The resolution is worst in the transition region $1.3 < \abseta < 1.6$.

\subsection{Additional offline tau calibration corrections and systematic uncertainties}\label{sec:offlineres}
The systematic uncertainties on the tau energy scale are evaluated with two complementary methods. The 
{\it deconvolution method} gives access to uncertainties on both the absolute TES (differences between reconstructed 
and true visible energy) and the modelling (differences between data and simulation) and is based on 
dedicated measurements (such as test beam data and low-luminosity runs) and simulation. The 
{\it in-situ method} only tests the modelling and uses collision data with typical 2012 LHC run 
conditions. Both methods are also able to provide small additional data-driven corrections albeit only 
inclusively in \et\ and \abseta\ due to the limited statistical power of the dataset. They thus depend 
on the first calibration step explained in the previous section to remove any significant TES dependencies on 
kinematics or pile-up.

The deconvolution method is almost identical to the method employed to measure the jet energy scale 
for ATLAS in 2010~\cite{jes} and is only briefly described here. The central idea is to 
decompose each tau lepton into its decay products and to combine the calorimeter responses according to the branching ratios 
of tau leptons to the various hadronic final states. The response to charged hadrons is estimated from different sources
depending on the momentum and pseudorapidity; in-situ $E/p$ measurements are used at low momentum,
combined test beam measurements are used at high momentum in the central region ($\abseta<0.8$),
and simulation is used otherwise (here, the uncertainty is estimated using events simulated using different hadronic shower models). 
The response to electromagnetic showers was studied in $Z \to ee$ decays and is used for neutral pions. 
Pseudo-experiments are used to propagate the single-particle response uncertainties to the reconstructed 
hadronically decaying tau lepton. In each pseudo-experiment, the tau decay product energies are 
varied randomly using Gaussian distributions centred on the observed ratio of the response in data and 
simulation and with a width corresponding to the statistical uncertainty, and Gaussian distributions centred at unity 
and with widths given by each systematic uncertainty. These distributions depend on particle type, energy and pseudorapidity. 
The TES shift for a single pseudo-experiment is given by the mean of the energy ratio of the \tauhadvis to an identical 
pseudo-experiment in which only statistical uncertainties of the measurement are considered by Gaussian 
distributions centred at unity. The distribution of TES shifts for a large number of pseudo-experiments is 
fitted with a Gaussian function. The mean of the fit is the expected scale shift between data and simulation, and its 
standard deviation the contribution to the TES uncertainty. 

Additional contributions considered
are uncertainties due to the detector modelling in the simulation,
the underlying event, the effect of pile-up, the non-closure of 
the calibration method (meaning the difference between the reconstructed and the true \tauhadvis energy, when applying the calibration to the same sample it was derived from) and the hadronic-shower model, as shown in Table~\ref{tab:tes_deconv}. 
The total TES uncertainty for $\ET>20$ \GeV\ and $\abseta<2.5$ is between 2\% and 3\% for \tauhadvis with one track 
and between 2\% and 4\% for \tauhadvis with more tracks, depending on \ET\ and \abseta. A TES shift of 
1\% is observed with no significant dependence on \ET\ or \abseta\ and a trend towards slightly higher 
values for 3-track \tauhadvis candidates. 
The shift is 
dominantly due to $E/p$ response differences between data and simulation.

\begin{table}[htp]
\begin{center}
\begin{tabular}{lc}
\hline\hline\noalign{\smallskip}
Source & Uncertainty [\%] \\
\noalign{\smallskip}\hline\noalign{\smallskip}
Response       & 1.2--2.5 \\
Detector model & 0.3--2.5 \\
UE             & 0.2--2.4 \\
Pile-up        & 0.5--2.0 \\
Non-closure    & 0.5--1.2 \\
Shower model   & 0.0--2.0 \\
\noalign{\smallskip}\hline\noalign{\smallskip}
Total          & 1.8--3.9 \\
\hline\hline\noalign{\smallskip}
\end{tabular}
\end{center}
\caption{
Systematic uncertainties on the tau energy scale estimated using the deconvolution method. In general, the values depend on 
\ET, \abseta $~$ and the number of associated tracks. The range of  values for $\ET>20$ \GeV\ is shown.
\label{tab:tes_deconv}
}
\end{table}

\begin{figure}[!bhp]
\begin{center}
\subfigure[]{
        \includegraphics[width=0.47\textwidth, height=6.6cm]{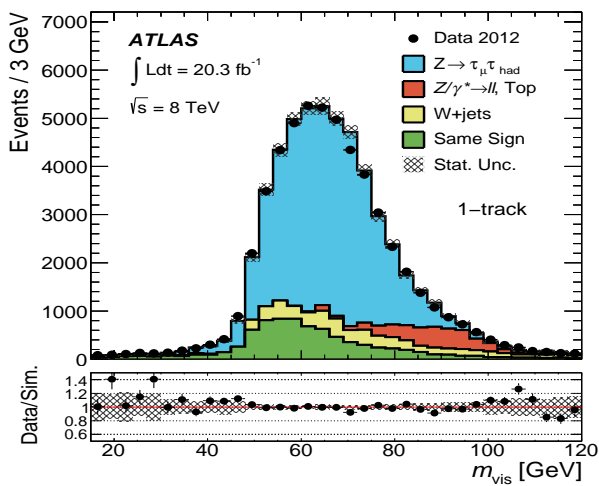}
}
\subfigure[]{
        \includegraphics[width=0.47\textwidth, height=6.6cm]{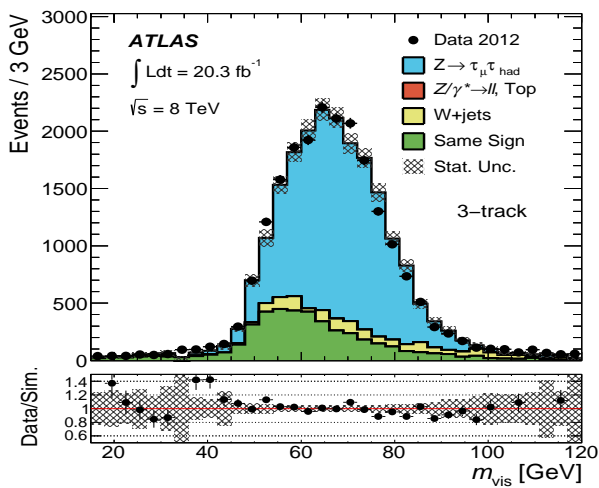}
}
\caption{
The \mvis\ distribution used for the in-situ offline TES measurement. Shown is the comparison 
between data and simulation for \tauhadvis with (a) one or (b) three associated tracks.
}\label{fig:insitu}
\end{center}
\end{figure}

The in-situ method is based on the fact that the distribution of the reconstructed visible mass \mvis\ in $Z \to \tau\tau$ 
events where one tau decays hadronically and the other to a muon plus neutrinos can be used to measure a TES shift between 
data and simulation and its uncertainty. Here, \mvis\ is defined as the invariant mass of the \tauhadvis and the muon.
The muon momentum scale is measured independently with high precision. The TES shift 
$\alpha$ is determined by introducing an energy shift 
$E_{\mathrm{T}} \to (1+\alpha) E_{\mathrm{T}} $
for \tauhadvis objects
and finding the value $\alpha$ for which the \mvis\ peak position in data and simulation agrees. A fifth-order polynomial fit is used to estimate
the \mvis\ peak position as simulation studies show that this gives both the highest sensitivity and robustness.
For small values of $\alpha$, the \mvis\ peak position depends linearly on \ET.

The results are based on collision data recorded by the ATLAS detector in 2012 using a muon trigger threshold of 24 \GeV. The event selection 
is similar to the one used by the $Z \to \tau\tau$ tag-and-probe studies described in Sect.~\ref{sec:effoffline} 
with the following differences: 
the \tauhadvis candidates are required to have $\ET > 20$\,\GeV\ and to satisfy {\it medium} \tauid criteria. No selection requirement is applied to
\mvis, and a looser $\cos \Delta \phi>-0.5$ requirement is made. Additionally, a pseudorapidity difference 
between the \tauhadvis and the muon smaller than 1.5 as well as $E^\tau_{\mathrm{T,vis}}-\ET^\mu > -15$\,\GeV\ is required.
The motivation for the differences is that this measurement requires a highly pure sample of hadronically 
decaying tau leptons after applying \tauid while the priority of the efficiency measurement is to obtain 
a largely unbiased sample before applying any identification requirements.

The background contributions are estimated in the same way as described in Sect.~\ref{sec:triggerperf}.
The dominant systematic uncertainties of the in-situ measurement are estimated using 
pseudo-experiments and are due to a potential bias of the fit, missing transverse 
momentum resolution and scale, muon momentum resolution, muon trigger efficiency and the 
normalization of the multi-jet background. They are summarized in Table~\ref{tab:tes_insitu}.

\begin{table}[htp]
\begin{center}
\begin{tabular}{lc}
\hline\hline\noalign{\smallskip}
Source & Uncertainty [\%] \\
\noalign{\smallskip}\hline\noalign{\smallskip}
Fit bias             & 0.5     \\
\met{} resolution    & 0.2     \\
\met{} scale         & 0.1     \\
$\pt^\mu$ resolution & 0.1--0.3 \\
Trigger              & 0.1     \\
Jet background       & 0.1--0.3 \\
\noalign{\smallskip}\hline\noalign{\smallskip}
Total                & 0.6--0.7 \\
\hline\hline\noalign{\smallskip}
\end{tabular}
\end{center}
\caption{
Dominant systematic uncertainties on the tau energy scale estimated using the in-situ method. In general, the values depend on 
the number of associated tracks. All other systematic uncertainties are smaller than 0.1\%.
\label{tab:tes_insitu}
}
\end{table}

The measured TES shift is  $\alpha =  0.8\% ~\pm ~1.3$\% (stat) $\pm ~0.6$\% (syst) 
and $\alpha=1.1\% ~\pm ~1.4$\% (stat) $\pm ~0.7$\% (syst) for \tauhadvis with one or three associated tracks 
respectively. No significant dependence on $\eta$ or pile-up conditions is observed. 
The corrections are {\it positive}, i.e. the momentum of \tauhadvis in data has to be scaled 
up in order to yield agreement (on average) with simulation, and are in agreement 
with the bias observed in data using the deconvolution method.
The resulting \mvis\ distribution for data and simulation is shown in Fig.~\ref{fig:insitu} 
before applying any correction (i.e., $\alpha=0$).
The uncertainties given above only account for differences between data and simulation and 
not in the absolute TES. For the latter, uncertainties due to non-closure and pile-up 
conditions estimated with the 
deconvolution method have to be added in quadrature to the systematic uncertainties given above.

\subsection{Trigger \tauhadvis energy calibration and resolution}

As described in Sect.~\ref{sec:triggerreco}, reconstructed \tauhadvis candidates at both L1 and L2 use a dedicated energy reconstruction algorithm 
which differs from the offline \tauhadvis energy reconstruction and calibration, while at the EF, the same algorithm is used.
In this section, comparisons of the online energy calibrations between data and simulation are shown.

\begin{figure}
\begin{center}
\subfigure[]{
\includegraphics[width=0.44\textwidth, height=6.6cm]{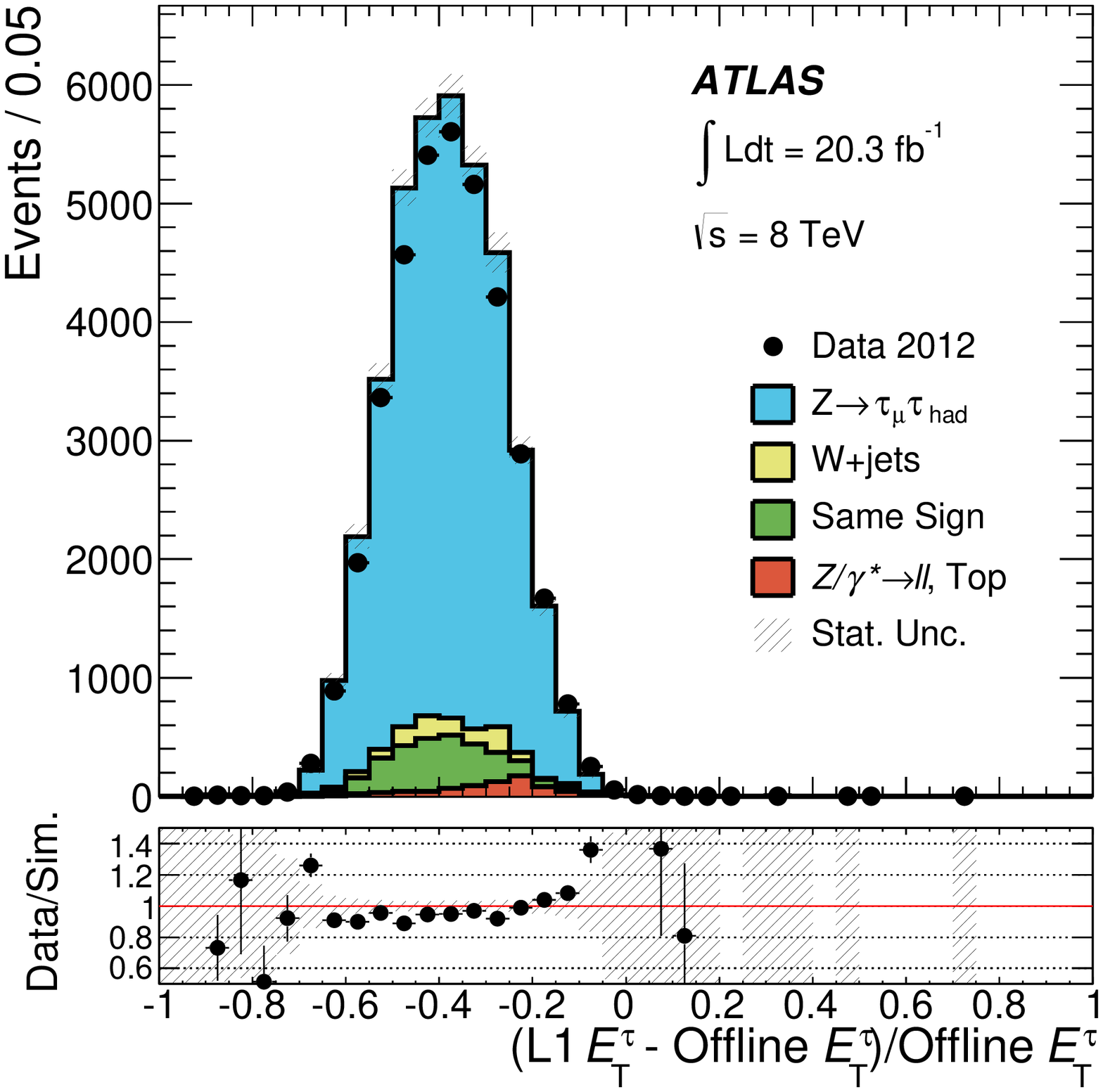}
}
\subfigure[]{
\includegraphics[width=0.44\textwidth, height=6.6cm]{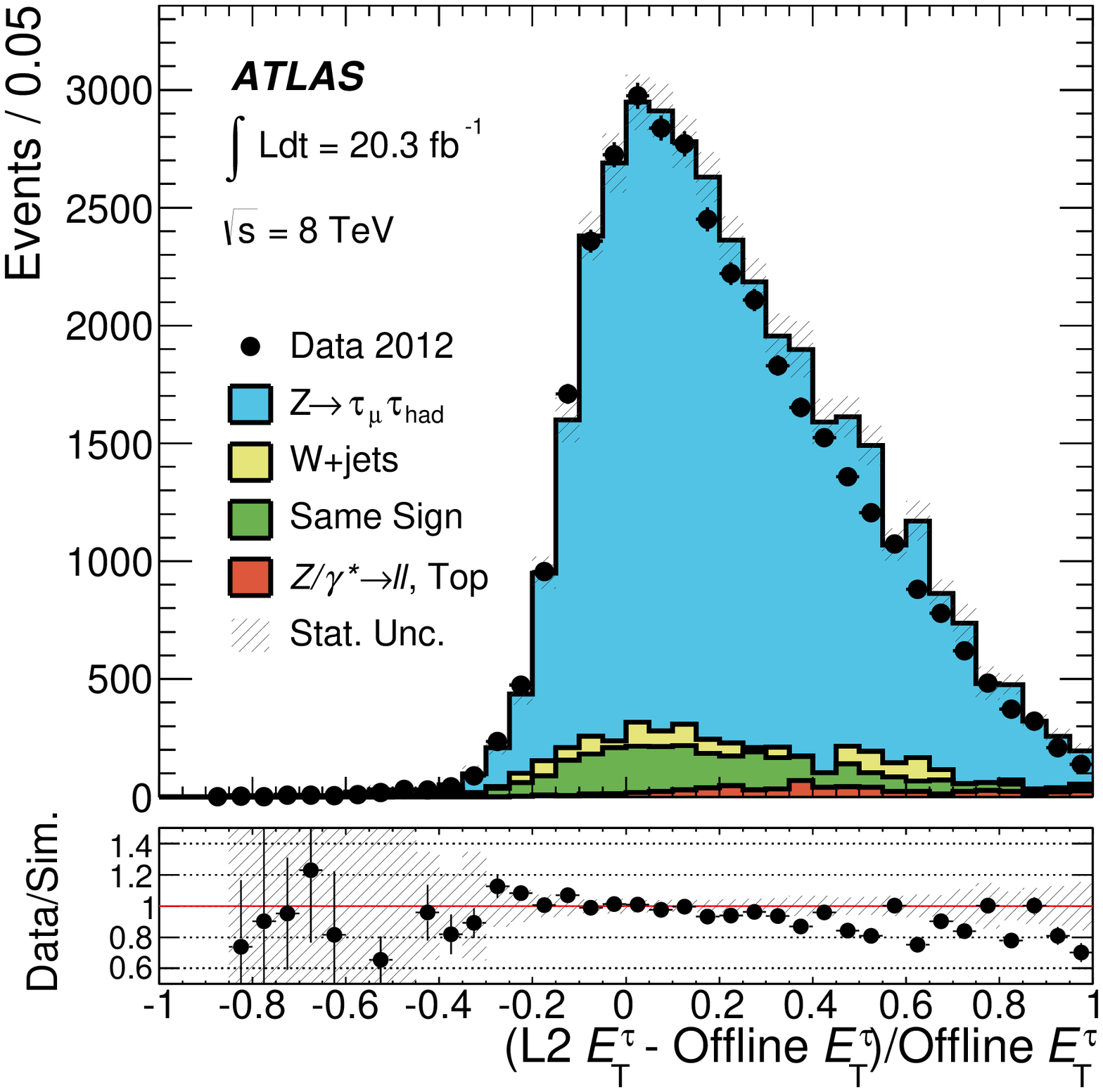}
}
\subfigure[]{
\includegraphics[width=0.44\textwidth, height=6.6cm]{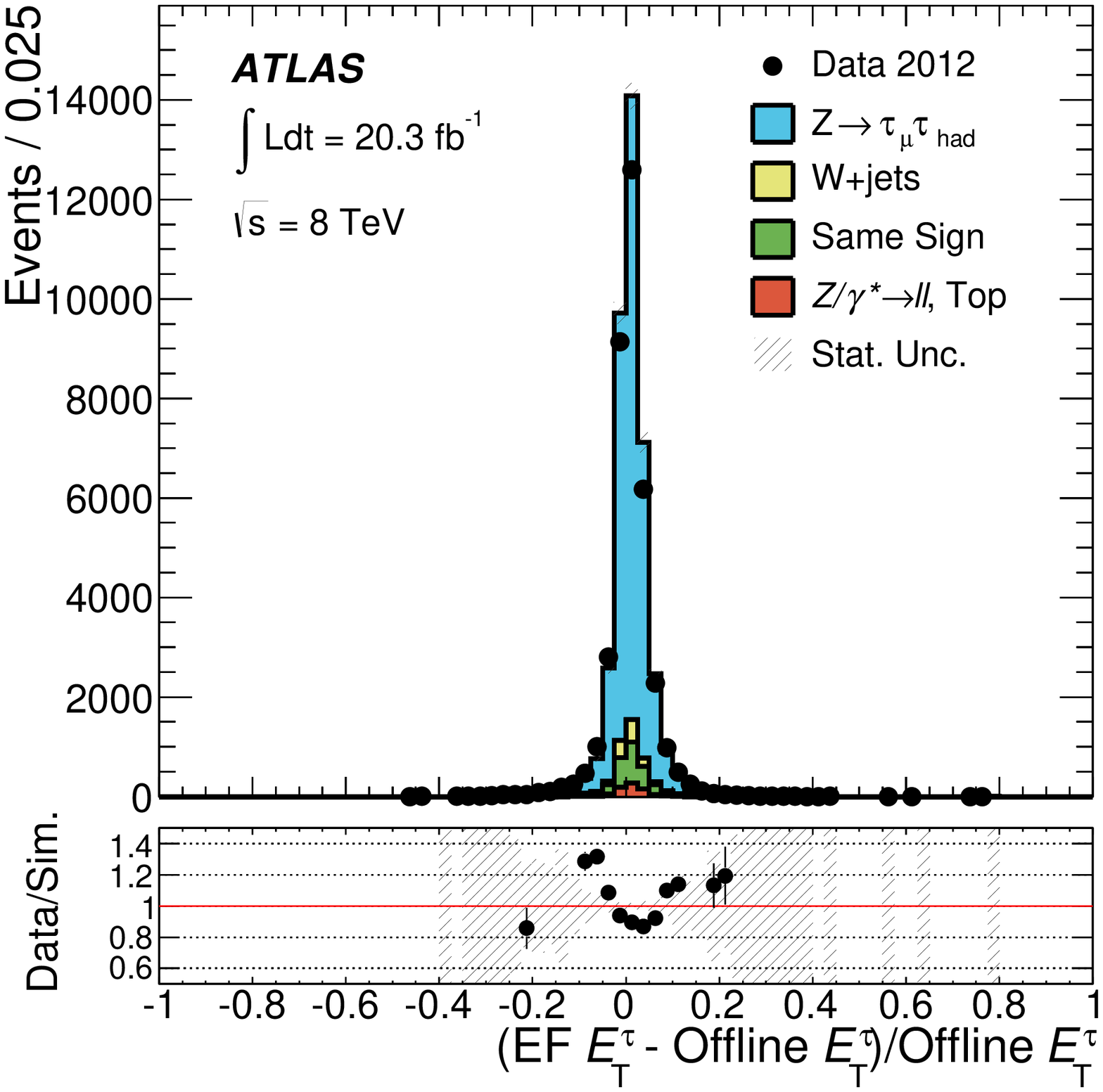}
}
\caption{
The measured tau trigger transverse energy resolution for the offline \tauhadvis candidates passing {\it medium} \tauid
at (a) L1, (b) L2 and (c) the EF.
The grey hashed area reflects the statistical uncertainties on the sum of the expected signal and background. 
}
\label{fig:trigreso}
\end{center}

\end{figure}

The measured transverse energy resolution for offline \tauhadvis candidates passing {\it medium} \tauid is shown in Fig.~\ref{fig:trigreso}
at all three trigger levels. This measurement is carried out using the same methodology as described in the previous section. 
The reconstructed energy at L1 is underestimated since at this level calorimeter energies are calibrated 
at the EM scale. The overestimation seen at L2 is due to the clustering algorithm used at L2, which does not implement the same
noise suppression scheme as offline. At the EF, the energy reconstruction is almost identical to the offline case. 
The slight difference with respect to the offline energy resolution is mainly due to the pile-up corrections, which are only
applied offline.  Some discrepancies can be seen between the resolutions measured in data and in simulation.  This reinforces
the importance of having a trigger efficiency measurement performed directly in data as a function of the offline \tauhadvis \pt{}, as
presented in section~\ref{sec:triggerperf}.

\section{Summary and conclusions}\label{sec:conclusions}

The algorithms developed in the ATLAS experiment at the LHC for tau identification and tau energy calibration are described, along with
their optimization and the associated procedures to mitigate the effects of pile-up.  These algorithms were employed in the dataset corresponding to
20.3 fb$^{-1}$ of $\sqrt{\mathrm{s}} = 8$ TeV $pp$ collisions.
The performance of the tau algorithms have helped to fulfil a variety of physics searches and measurements with hadronically decaying tau leptons, 
an important part of the ATLAS physics program.
The performance of trigger and offline tau identification and calibration is measured, in most cases using $Z \to \tau\tau$ tag-and-probe measurements. 
The uncertainties on the offline tau identification efficiency measurement are dependent on the working point and are about 
(2--3)\% for \tauhadvis with one associated track, and (4--5)\% for the case of three associated tracks, inclusive in $\eta$ and 
for a visible transverse momentum greater than $20$ \GeV.
A precision of (2--8)\% for the tau trigger identification efficiency is measured for hadronic tau decays selected by offline algorithms, depending
on the transverse energy.
Stability of all algorithms with respect to the pile-up conditions is observed. 
The reconstructed tau energy scale is measured with a precision of about (2--4)\% depending on transverse energy and pseudorapidity, 
using either a method based on estimating and deconvolving the response uncertainties of 
the hadronic tau decay products or a direct measurement of the $Z \to \tau\tau$ visible mass using collision data recorded in 2012.



\begin{acknowledgements}

We thank CERN for the very successful operation of the LHC, as well as the
support staff from our institutions without whom ATLAS could not be
operated efficiently.

We acknowledge the support of ANPCyT, Argentina; YerPhI, Armenia; ARC,
Australia; BMWFW and FWF, Austria; ANAS, Azerbaijan; SSTC, Belarus; CNPq and FAPESP,
Brazil; NSERC, NRC and CFI, Canada; CERN; CONICYT, Chile; CAS, MOST and NSFC,
China; COLCIENCIAS, Colombia; MSMT CR, MPO CR and VSC CR, Czech Republic;
DNRF, DNSRC and Lundbeck Foundation, Denmark; EPLANET, ERC and NSRF, European Union;
IN2P3-CNRS, CEA-DSM/IRFU, France; GNSF, Georgia; BMBF, DFG, HGF, MPG and AvH
Foundation, Germany; GSRT and NSRF, Greece; RGC, Hong Kong SAR, China; ISF, MINERVA, GIF, I-CORE and Benoziyo Center, Israel; INFN, Italy; MEXT and JSPS, Japan; CNRST, Morocco; FOM and NWO, Netherlands; BRF and RCN, Norway; MNiSW and NCN, Poland; GRICES and FCT, Portugal; MNE/IFA, Romania; MES of Russia and NRC KI, Russian Federation; JINR; MSTD,
Serbia; MSSR, Slovakia; ARRS and MIZ\v{S}, Slovenia; DST/NRF, South Africa;
MINECO, Spain; SRC and Wallenberg Foundation, Sweden; SER, SNSF and Cantons of
Bern and Geneva, Switzerland; NSC, Taiwan; TAEK, Turkey; STFC, the Royal
Society and Leverhulme Trust, United Kingdom; DOE and NSF, United States of
America.

The crucial computing support from all WLCG partners is acknowledged
gratefully, in particular from CERN and the ATLAS Tier-1 facilities at
TRIUMF (Canada), NDGF (Denmark, Norway, Sweden), CC-IN2P3 (France),
KIT/GridKA (Germany), INFN-CNAF (Italy), NL-T1 (Netherlands), PIC (Spain),
ASGC (Taiwan), RAL (UK) and BNL (USA) and in the Tier-2 facilities
worldwide.

\end{acknowledgements}

\bibliography{}   

\providecommand{\href}[2]{#2}\begingroup\raggedright\begin{thebibliography}{10}

\bibitem{PDG}
   K.A.~Olive {\em et.~al.}, (Particle Data Group), 
   Chin. Phys. C {\bf 38}, 090001 (2014)

\bibitem{Aad:2012vip}
  ATLAS Collaboration, 
  Eur. Phys. J. C {\bf 73}, 2328 (2013) [\href{http://xxx.lanl.gov/abs/1211.7205}{{\tt arXiv:1211.7205}}]

\bibitem{Aad:2012mza}
  ATLAS Collaboration, 
  Phys. Lett. B {\bf 717}, 89--108 (2012) [\href{http://xxx.lanl.gov/abs/1205.2067}{{\tt arXiv:1205.2067}}]

\bibitem{Aad:2012cia}
  ATLAS Collaboration, 
  Eur. Phys. J. C {\bf 72}, 2062 (2012) [\href{http://xxx.lanl.gov/abs/1204.6720}{{\tt arXiv:1204.6720}}]

\bibitem{Aad:2011fu}
  ATLAS Collaboration, 
  Phys. Lett. B {\bf 706}, 276--294 (2012) [\href{http://xxx.lanl.gov/abs/1108.4101}{{\tt arXiv:1108.4101}}]

\bibitem{Aad:2011kt}
  ATLAS Collaboration, 
  Phys. Rev. D {\bf 84}, 112006 (2011) [\href{http://xxx.lanl.gov/abs/1108.2016}{{\tt arXiv:1108.2016}}]

\bibitem{Aad:2012mea}
  ATLAS Collaboration, 
  JHEP {\bf 09}, 070 (2012) [\href{http://xxx.lanl.gov/abs/1206.5971}{{\tt arXiv:1206.5971}}]

\bibitem{Aad:2012rjx}
  ATLAS Collaboration, 
  JHEP {\bf 03}, 076 (2013) [\href{http://xxx.lanl.gov/abs/1212.3572}{{\tt arXiv:1212.3572}}]

\bibitem{Aad:2012tj}
  ATLAS Collaboration, 
  JHEP {\bf 06}, 039 (2012) [\href{http://xxx.lanl.gov/abs/1204.2760}{{\tt arXiv:1204.2760}}]

\bibitem{Aad:2012cfr}
  ATLAS Collaboration, 
  JHEP {\bf 02}, 095 (2013) [\href{http://xxx.lanl.gov/abs/1211.6956}{{\tt arXiv:1211.6956}}]

\bibitem{Aad:2012ypy}
  ATLAS Collaboration, 
  Phys. Lett. B {\bf 723}, 15--32 (2013) [\href{http://xxx.lanl.gov/abs/1212.1272}{{\tt arXiv:1212.1272}}]

\bibitem{Aad:2014mra}
  ATLAS Collaboration, 
  {\em Submitted to JHEP} [\href{http://xxx.lanl.gov/abs/1407.0603}{{\tt arXiv:1407.0603}}]

\bibitem{Aad:2014yka}
  ATLAS Collaboration, 
  {\em Submitted to JHEP} [\href{http://xxx.lanl.gov/abs/1407.0350}{{\tt arXiv:1407.0350}}]

\bibitem{Aad:2012gm}
  ATLAS Collaboration, 
  Phys. Lett. B {\bf 719}, 242--260 (2013) [\href{http://xxx.lanl.gov/abs/1210.6604}{{\tt arXiv:1210.6604}}]

\bibitem{ATLAS:2013oea}
  ATLAS Collaboration, 
  JHEP {\bf 06}, 033 (2013) [\href{http://xxx.lanl.gov/abs/1303.0526}{{\tt arXiv:1303.0526}}]

\bibitem{dt}
  L.~Breiman, J.~Friedman, R.~Olshen, and C.~Stone, 
  {\it Classification and Regression Trees},
  \newblock Chapman \& Hall, New York, (1984)

\bibitem{adaboost}
  Y.~Freund and R.~E. Schapire, 
  J. Comput. Syst. Sci. {\bf 55} (1997)

\bibitem{LCCalibration}
  T. Barillari et al., 
  ATL-LARG-PUB-2009-001-2 (2009),
  [\href{http://cds.cern.ch/record/1112035}{{\tt http://cds.cern.ch/record/1112035}}]

\bibitem{cscbook}
  ATLAS Collaboration, 
  SLAC-R-980, CERN-OPEN-2008-020 (2008), [\href{http://xxx.lanl.gov/abs/0901.0512}{{\tt arXiv:0901.0512}}]

\bibitem{atlas_det}
  ATLAS Collaboration,
  JINST {\bf 3}, S08003 (2008)

\bibitem{lhc}
  L.~Evans and P.~Bryant, 
  JINST {\bf 3}, S08001 (2008)

\bibitem{egammaPaper1}
  ATLAS Collaboration, 
  Eur. Phys. J. C {\bf 74}, 2941 (2014)
  [\href{http://xxx.lanl.gov/abs/1404.2240}{{\tt arXiv:1404.2240}}]

\bibitem{egammaPaper2}
  ATLAS Collaboration, 
  {\em Submitted to EPJC} [\href{http://xxx.lanl.gov/abs/1407.5063}{{\tt arXiv:1407.5063}}]

\bibitem{muonPaper}
  ATLAS Collaboration, 
  {\em Submitted to EPJC} [\href{http://xxx.lanl.gov/abs/1407.3935}{{\tt arXiv:1407.3935}}]

\bibitem{antikt}
  M.~Cacciari, G.~P. Salam, and G.~Soyez,
  JHEP {\bf 04}, 063 (2008) [\href{http://xxx.lanl.gov/abs/0802.1189}{{\tt arXiv:0802.1189}}]

\bibitem{TopoClusters}
  W.~Lampl {\em et.~al.},
  ATL-LARG-PUB-2008-002 (2008),
  [\href{http://cds.cern.ch/record/1099735}{{\tt http://cds.cern.ch/record/1099735}}]

\bibitem{met}
  ATLAS Collaboration, 
  Eur. Phys. J. {\bf C72}, 1844 (2012)

\bibitem{Alpgen}
  M.~L. Mangano, M.~Moretti, F.~Piccinini, R.~Pittau and A.~D. Polosa, 
  JHEP {\bf 07}, 001 (2003) [\href{http://xxx.lanl.gov/abs/hep-ph/0206293}{{\tt hep-ph/0206293}}]

\bibitem{Herwig}
  G.~Corcella, I.~G. Knowles, G.~Marchesini, S.~Moretti, K.~Odagiri, P.~Richardson, M.~H. Seymour and B.~R. Webber,
  JHEP {\bf 01}, 010 (2001)  [\href{http://xxx.lanl.gov/abs/hep-ph/0011363}{{\tt hep-ph/0011363}}], [\href{http://xxx.lanl.gov/abs/hep-ph/0210213}{{\tt hep-ph/0210213}}]


\bibitem{Pythia6}
  T.~Sjostrand, S.~Mrenna, and P.~Skands, 
  JHEP {\bf 05}, 026 (2006) [\href{http://xxx.lanl.gov/abs/hep-ph/0603175}{{\tt hep-ph/0603175}}]

\bibitem{Pythia8}
  T.~Sjostrand, S.~Mrenna, and P.~Z. Skands, 
  Comput. Phys. Commun. {\bf 178}, 852--867 (2008)  [\href{http://xxx.lanl.gov/abs/0710.3820}{{\tt arXiv:0710.3820}}]

\bibitem{MCNLO}
  S.~Frixione and B.~R. Webber, 
  JHEP {\bf 06}, 029 (2002)

\bibitem{AcerMC}
  B. P. Kersevan and E. Richter-W\c{a}s, 
  [\href{http://xxx.lanl.gov/abs/0405247}{{\tt hep-ph/0405247}}]

\bibitem{Tauola}
  Z.~Was and P.~Golonka, 
  Nucl. Phys. Proc. Suppl. {\bf 144}, 88--94 (2005) [\href{http://xxx.lanl.gov/abs/0411377}{{\tt hep-ph/0411377}}]

\bibitem{Photos}
  E.~Barberio, B.~van Eijk, and Z.~Was, 
  Comput. Phys. Commun. {\bf 66}, 115--128 (1991)

\bibitem{ct6}
  J.~Pumplin {\em et.~al.}, 
  JHEP {\bf 07}, 012 (2002) [\href{http://xxx.lanl.gov/abs/hep-ph/0201195}{{\tt hep-ph/0201195}}]

\bibitem{ct10}
  H.-L. Lai {\em et.~al.}, 
  Phys. Rev. D {\bf 82}, 074024 (2010) [\href{http://xxx.lanl.gov/abs/1007.2241}{{\tt arXiv:1007.2241}}]

\bibitem{PUB-2012-003}
  ATLAS Collaboration,
  ATL-PHYS-PUB-2012-003 (2012),
  [\href{http://cds.cern.ch/record/1474107}{{\tt http://cds.cern.ch/record/1474107}}]

\bibitem{PUB-2011-008}
  ATLAS Collaboration,
  ATL-PHYS-PUB-2011-008 (2011),
  [\href{http://cds.cern.ch/record/1345343}{{\tt http://cds.cern.ch/record/1345343}}]

\bibitem{p2011tune}
  P.~Z. Skands, 
  Phys. Rev. D {\bf 82}, 074018 (2010) [\href{http://xxx.lanl.gov/abs/1005.3457}{{\tt arXiv:1005.3457}}]

\bibitem{PUB-2011-009}
  ATLAS Collaboration,
  ATL-PHYS-PUB-2011-009 (2011),
  [\href{http://cds.cern.ch/record/1363300}{{\tt http://cds.cern.ch/record/1363300}}]

\bibitem{Atlassim}
  ATLAS Collaboration, 
  Eur. Phys. J. C {\bf 70}, 823 (2010) [\href{http://arxiv.org/abs/1005.4568}{{\tt arXiv:1005.4568}}]

\bibitem{geant}
  GEANT4 Collaboration, S.~Agostinelli {\em et.~al.}, 
  Nucl. Instr. Methods Phys. Res. A {\bf 506}, 250--303 (2003)

\bibitem{Folger:2003sb}
  G.~Folger and J.~Wellisch, 
  {\em eConf} {\bf C0303241} MOMT007, (2003) [\href{http://xxx.lanl.gov/abs/nucl-th/0306007}{{\tt nucl-th/0306007}}]


\bibitem{Bertini}
  H.~W. Bertini, 
  Phys. Rev. {\bf 188}, 1711--1730 (1969)

\bibitem{FTF}
  B.~Andersson, G.~Gustafson, and B.~Nilsson-Almqvist,
  Nucl. Phys. B {\bf 281}, no.~1-2 289--309 (1987)

\bibitem{JVF}
  ATLAS Collaboration,
  ATLAS-CONF-2014-018 (2014),
  [\href{http://cds.cern.ch/record/1700870}{{\tt http://cds.cern.ch/record/1700870}}]

\bibitem{Aad:2012xs}
  ATLAS Collaboration, 
  Eur. Phys. J. C {\bf 72}, 1849 (2012) [\href{http://xxx.lanl.gov/abs/1110.1530}{{\tt arXiv:1110.1530}}]

\bibitem{jes}
  ATLAS Collaboration, 
  Eur. Phys. J. C {\bf 73}, 2304 (2013) [\href{http://xxx.lanl.gov/abs/1112.6426}{{\tt arXiv:1112.6426}}]

\end{thebibliography}\endgroup

\clearpage

\onecolumn
\begin{center}
\begin{flushleft}
{\Large The ATLAS Collaboration}

\bigskip

G.~Aad$^{\rm 85}$,
B.~Abbott$^{\rm 113}$,
J.~Abdallah$^{\rm 152}$,
S.~Abdel~Khalek$^{\rm 117}$,
O.~Abdinov$^{\rm 11}$,
R.~Aben$^{\rm 107}$,
B.~Abi$^{\rm 114}$,
M.~Abolins$^{\rm 90}$,
O.S.~AbouZeid$^{\rm 159}$,
H.~Abramowicz$^{\rm 154}$
% ...
\bigskip
\\
% ...
$^{11}$ Institute of Physics, Azerbaijan Academy of Sciences, Baku, Azerbaijan\\
% ...
$^{85}$ CPPM, Aix-Marseille Universit{\'e} and CNRS/IN2P3, Marseille, France\\
% ...
$^{90}$ Department of Physics and Astronomy, Michigan State University, East Lansing MI, United States of America\\
% ...
$^{107}$ Nikhef National Institute for Subatomic Physics and University of Amsterdam, Amsterdam, Netherlands\\
% ...
$^{113}$ Homer L. Dodge Department of Physics and Astronomy, University of Oklahoma, Norman OK, United States of America\\
$^{114}$ Department of Physics, Oklahoma State University, Stillwater OK, United States of America\\
% ...
$^{117}$ LAL, Universit{\'e} Paris-Sud and CNRS/IN2P3, Orsay, France\\
% ...
$^{152}$ Institute of Physics, Academia Sinica, Taipei, Taiwan\\
% ...
$^{154}$ Raymond and Beverly Sackler School of Physics and Astronomy, Tel Aviv University, Tel Aviv, Israel\\
% ...
$^{159}$ Department of Physics, University of Toronto, Toronto ON, Canada\\
% ...
\end{flushleft}


\end{center}

\end{document}
